\documentclass[12pt,a4paper]{article}
\usepackage{graphicx}
\usepackage[cp1251]{inputenc}
\usepackage[english]{babel}
\usepackage{amssymb, amsfonts, amsmath}
\usepackage{amscd}
\begin{document}

\date{}

\title{EXACT SOLUTIONS OF MAXWELL VACUUM EQUATIONS IN PETROV HOMOGENEOUS NON-NULL SPACES}

\author{V. V. Obukhov}

\maketitle

\noindent
Institute of Scietific Research and Development,
Tomsk State Pedagogical University (TSPU). Tomsk State  Pedagogical University, 60 Kievskaya St., Tomsk, 634041, Russia; \\ \quad

\noindent
Laboratory for Theoretical Cosmology, International Center of Gravity and Cosmos, Tomsk State University of Control Systems and Radio Electronics (TUSUR), 36, Lenin Avenue, Tomsk, 634050, Russia

\quad

\noindent
Correspondence: obukhov@tspu.ru;

\quad

\noindent
Keywords: theory of symmetry, continuous groups of transformations, homogeneous spaces, Maxwell equation.

\section{INTRODUCTION}

Of particular interest in the theory of gravity and electromagnetism are spaces and fields endowed with symmetry. The symmetry of space is manifested through the symmetry of geodesic lines and is defined by vector or tensor Killing fields. In addition, this symmetry is manifested in physical fields through the motion integrals of such equations as the Hamilton-Jacobi, Klein-Gordon-Fock, Dirac-Fock, etc. equations.




In gravity theory, the most interesting spaces are those with Lorentz signature that admit three Killing fields. These spaces include two types of spaces: Stackel spaces (see \cite{1}–\cite{9}) and Petrov spaces (see \cite{10}–\cite{12}). Both types of the spaces have complete sets consisting of three Killing fields. For Stackel spaces, the complete sets consist of mutually commuting Killing tensor fields of the first rank (vector fields) and/or of the second rank. For Petrov spaces, the complete sets consist of three Killing vector fields. Such sets allow one to introduce a privileged coordinate system in which the components of the metric tensor contain functions, each of which depends on only one of the variables.

This makes it possible to solve the separation of variables  problem  (including in the presence of physical fields) in  the field equations and in the motion equations for a single test particle.

The above-mentioned feature of this type spaces allows to reduce these equations to systems of ordinary differential equations. All the exact solutions of motion equations known to date were found thanks to precisely this circumstance. For Stackel spaces (see \cite{13}-\cite{18}). For Petrov spaces see \cite{10}, \cite{12}, \cite{19}-\cite{25}. Among these solutions are such important solutions as the solutions \cite{26}-\cite{33}

Stackel spaces continue to attract the attention of researchers, since they are used as a basis for constructing models of real processes occurring in the presence of a gravitational field. Among them, metrics with spherical or axial symmetry are most often considered, including in multidimensional theories of gravity. As an example, we can cite the papers \cite{34}-\cite{40}. As examples of the study of Petrov spaces in cosmology, one can cite recent paperes:  \cite{401} (where the Hamiltonian approach to cosmological models according to Bianchi is developed)
and \cite{402} (authors claim to have obtained an ideal characterization of spatially homogeneous cosmologies).

One of the most important examples of metrics related to the class of Stackel spaces are the De Sitter, Anti De Sitter, Friedman and some other metrics that underlie the vast majority of cosmological models, including models of the modified theory of gravity. A huge number of articles are devoted to these models (see, for example, \cite{42}-\cite{56}).

Let us note that recently wave spaces have attracted considerable interest. The manyfold of such spaces intersects with the manyfold of null Stackel spaces, which have null Killing vector fields in  its complete sets.  The general form of the metric of an null Stackel space was first found by V.N. Shapovalov. Therefore in paper \cite{57} it was proposed to call null Stackel spaces with an null Killing vector fields as Shapovalov wave spaces. As an example of works devoted to various aspects of Shapovalov wave spaces, one can cite papers \cite{58}-\cite{62}.

In Petrov spaces, another method of exact integration of the equations of motion of a test particle is used. It is based on the use of a non-commutative group of motions of the space. Therefore, it is called the method of non-commutative integration. (see article \cite{19}).
Note that a large number of articles are devoted to the consideration of other problems in Petrov spaces
(see articles \cite{63}-\cite{69}), also to the classification of homogeneous Stackel spaces (see \cite{70}-\cite{72}).

 An important direction is the consideration of problems of symmetry theory in homogeneous spaces (see, for example, \cite{721} - \cite{723} ) too.

 \quad

 \noindent

In conclusion of this section, let's note the following.

\noindent
The results obtained after applying the theory of symmetry (including the theory of continuous groups of transformations) to the equations of mathematical physics make it possible to list all nonequivalent solutions of these equations that are invariant with respect to the corresponding symmetry operators. This is the purpose of the classification.
To date, classifications have been constructed (completely or partially) for the Laplace equations, d'Alembert equations, for single-particle equations of motion, etc.
After the appearance of the General relativity, the most important direction of classification became the classification of Riemannian and pseudo-Riemannian spaces with a given symmetry (the problem has been completely solved for Stackel spaces and for the homogeneous Petrov spaces), as well as the classification of mathematical physics equations solutions in these spaces. In addition, Einstein equations were added to the number of basic equations of mathematical physics. A large number of papers are devoted to solving classification problems in the General relativity.
From the point of view of mathematical physics, completely solved classification problems are of particular interest.

The range of such solved problems is expanded by the problem havs been solved in this article. A classification of exact solutions of Maxwell vacuum equations in homogeneous non-zero Petrov spaces with an invariant electromagnetic field is constructed. A complete list of such solutions is obtained.
 \quad

\section{NON-NULL PETROV SPACES}

Consider a Riemannian space $V_4$ with a Lorentz signature and with a group of motions $G_3(N)$ acting simply transitively on the hypersurface $V_3$. Here $N$ corresponds to the number of the group $G_3$ according to Bianchi classification. The hypersurface $V_3$ has the geometry of a three-dimensional homogeneous space $V_3(N)$. The space $V_4$ is called a homogeneous Petrov space and is denoted by $V_4(N)$. Petrov was the first who carried out a complete classification of the spaces $V_4(N)$ in papers \cite{10} \cite{11}.
In the case where the hypersurfaces of transitivity $V_{3}$ are non-null spaces with Euclidean signature, $V_{4}(N)$ are also denoted as homogeneous spaces of type $N$ according to the Bianchi classification.

Throughout the text, the following notations for indices are used:
\[i,j,k,r \div 1,2,3,0;\quad \alpha ,\beta ,\gamma ,\delta ... \div 1,2,3;\quad a,b,c,d\div 1,2,3.\]
The fourth coordinate of the holonomic coordinate system $\left\{u^i\right\}$ is supplied with the index $i=0$ ($u^0$), and the numbering of coordinate indices always starts with one and ends with zero (when indices $i,j,k,l$ are used). Greek letters denote the coordinate indices of the semi-geodetic coordinate system $\left\{u^{\alpha} \right\}$, related to the transitivity hypersurface of the group $G_{3} \left(N\right)$. Latin letters a, b, c, d denote the indices of the non-holonomic coordinate system associated with the group $G_{3} \left(N\right)$, as well as the indices of the structural constants and the numbers of the reper vectors.

\quad

The  Petrov method of classification of non-null spaces $V_{4}(N)$  is as follows:
\begin{enumerate}
\item
A semi-geodetic coordinate system, in which the metric tensor of the space $V_{4}(N)$ has the form:
\begin{equation} \label{1}
g^{ij} =\left(\begin{array}{cccc} {g^{\alpha \beta }(u^i) } & {} & {} & {0} \\ {} & {} & {} & {0} \\ {} & {} & {} & {0} \\ {0} & {0} & {0} & {-\varepsilon \tau^2(u^0)} \end{array}\right), \quad \det\left\|g_{\alpha \beta }\right \| =  \varepsilon G(u^0)^2, \quad \varepsilon = \pm 1,
\end{equation}
is introduced. Unless otherwise stated, we believe $\tau =1. $
Obviously, admissible coordinate transformations:
\begin{equation} \label{2}
\widetilde{u}^{\alpha} = \widetilde{u}^{\alpha}(u^{\beta}), \quad \widetilde{u}^0=\widetilde{u}^0(u^0)
\end{equation}
do not violated the form \eqref{1} of the metric tensor.
\item  The equations of the structure:
\begin{equation} \label{3}
\left [ X_a,X_b \right ] =C^c_{ab}X_c \quad (X_a = \xi_a^i p_i = \xi_a^i \partial_i)
\end{equation}
are integrated, and the Killing vector fields $\xi_a^i$ are found. Using admissible coordinate transformations \eqref{2}, arbitrariness in the obtained solutions is eliminated.

\noindent

\item
Using the functions ${\xi_a^\alpha}$, the components of the tensor  $g^{\alpha\beta}$ can be found by integrating the Killing equations:
\begin{equation} \label{4}
g^{\alpha\beta}_{,\gamma}\xi_{a}^{\gamma} = g^{\alpha\gamma}\xi^{\beta}_{a,\gamma} + g^{\beta\gamma}\xi^{\alpha}_{a,\gamma}.
\end{equation}
As is known, the metric tensor of the non-null homogeneous Petrov space can also be defined using a triad of vectors of the canonical frame  $\zeta^\alpha_{a}$,  satisfying the equations of the structure:
\begin{equation} \label{5}
\left [ Y_a,Y_b \right ] =C^c_{ab}Y_c, \quad (Y_a = \zeta_a^i p_i ).
\end{equation}
In the same coordinate system the vector fields $\xi_a^{\alpha}, \zeta_a^{\alpha} $ are interconnected by equations \begin{equation} \label{6}
\zeta_{a,\beta }^{\alpha } =\xi_{\beta }^b \xi_{b,\gamma }^{\alpha}\zeta_a^{\gamma} \Rightarrow
\xi_{a,\beta}^{\alpha} =\zeta_\beta^b \zeta_{b,\gamma}^{\alpha } \xi_a^{\gamma },
\end{equation}
where
$\xi^b_\alpha, \zeta^b_\alpha$ are covariant components of the Killing vectors and the canonical frame, respectively:
$$
\xi_{\beta }^a \xi_a^\alpha =\zeta_\beta^b \zeta_b^\alpha = \delta_\beta^\alpha, \quad \xi_{\alpha }^a \xi_b^\alpha =\zeta_\alpha^a \zeta_b^\alpha = \delta_b^a.
$$
(see \cite{78}, p. 484)
Therefore, the classification by Petrov method should be supplemented by a final stage:

\noindent
\item
\emph{Classification of all independent solutions of the equations \eqref{6}.}
\end{enumerate}
 This stage is performed in the paper \cite{12}. The integration of the equation of the structure \eqref{3} in the semi-geodetic coordinate system \eqref{1} has been carried out by A. Z. Petrov in the book \cite{10} (see further comments in the section 5 for the group of motions $G_3(VIII)$). The sets of vector fields obtained by Petrov were considered as sets of Killing vector fields $\xi^\alpha_a$ . However, they can also be considered as triads of vector fields $\zeta^\alpha_a$ of the canonical frame. In this case, one obtains an alternative classification method.
The metric tensor on the hypersurface $V_{3}(N)$ can be immediately found in the form:
\begin{equation}\label{7}
g_{\alpha\beta}= {\zeta_\alpha}^{a}{\zeta_\beta}^{b} \eta^{ab}(u^0) \Rightarrow  G = \zeta\eta \quad (\eta^2= \varepsilon\det\left\|\eta_{\alpha \beta }\right \|, \zeta= \det\left\|\zeta^{\alpha}_a\right \|).
\end{equation}

When studying physical processes occurring in Petrov spaces, it is necessary to know the integrals of motion of the geodesic equations.
In Petrov spaces, there are at least three integrals of motion linear in momenta. They are defined by Killing vector fields. Therefore, the classification of solutions to the system of equations \eqref{6} is the final stage in this method of classifying Petrov spaces. This stage will allow one to find the Killing vector fields $\xi^\alpha_a$ in the same coordinate system.

Let's note that the fields $\zeta^\alpha_a$ and $\xi^\alpha_a$ can be swapped. In this case, the coordinate transformation $ \quad \tilde{u}^\alpha =\tilde{u}^\alpha(u^\beta) $ should be carried out according to the formulas:
\begin{equation}\label{8}
{\tilde{\xi}}^\alpha_{a}(\tilde{u}) =\zeta_{a}^\alpha(\tilde{u}), \quad {\tilde{\zeta}}^\alpha_{a}(\tilde{u}) =\xi_{a}^\alpha(\tilde{u}).
\end{equation}
The new coordinate system $\left\{u^i\right\}$ can be found from the system of differential equations:
\begin{equation}\label{9}
\xi^\alpha_{a}(u) =\frac{\partial\tilde{u}^\alpha}{\partial u^\beta}\zeta_{a}^\beta(u).
\end{equation}
However, the  solution to this problem has no practical significance.

Obviously, when Petrov method is supplemented by the fourth stage, both methods give the same results presented in different coordinate systems. An alternative approach is more convenient when the classification results are used to integrate field equations with physical fields (for example, with electromagnetic fields) invariant with respect to the action of the group of motions $G_3(N)$.
In the present paper, this is demonstrated on example of Maxwell vacuum equations for an electromagnetic field invariant under the action of the group $G_3(N)$. The transition to a nonholonomic canonical frame allows one to reduce Maxwell equations to a system of algebraic equations. Note that the classification problem was considered within the framework of Petrov's method in \cite{73}, \cite{76}. Some of the solutions obtained there are presented in quadratures. Using the new method, one has obtained all solutions in explicit form. The method is common to all homogeneous nonnull Petrov spaces.

\quad

\noindent

The following problems are solved in the paper.

\quad

\noindent
1. All solutions of Maxwell vacuum equations have been obtained in explicit form and classified. The common solution scheme have been used for all groups of motions, which made it possible to significantly simplify the solution procedure and further simplify and systematize the previously obtained solutions.

\quad

\noindent
2. The classification of exact solutions of Maxwell vacuum equations for Petrov spaces with an unsolvable group of motions $G_3(VIII)$ has been supplemented by the classification for two new Petrov spaces.

\quad

\noindent
3. Section 6 provides a complete list of the solutions obtained.

\quad

\section{MAXWELL VACUUM EQUATIONS IN THE CANONICAL FRAME}

\quad

As it has been shown in the works \cite{25}, in the space $V_4(N)$ the Hamilton-Jacobi and Klein-Gordon-Fock equations for a charged particle have integrals of motion that are linear in momenta if the external electromagnetic field is invariant with respect to the group $G_r(N)$ acting simply transitively on the non-null hypersurface $V_3(N)$. Having chosen the gauge of the electromagnetic potential in the form:
$A_0=0,$ one obtains the following condition on the electromagnetic potential:
\begin{equation}\label{10}
(\xi_a^\alpha A_\alpha)_{,\beta}\xi_b^\beta=C^c_{ba}A_\alpha\xi_c^\alpha.
\end{equation}
Let us denote the nonholonomic components of the electromagnetic field vector in the frame $\xi_b^\alpha$ as following:
$$\textbf{A}_a =\xi_a^\alpha A_\alpha.$$
Then the equation \eqref{10} takes the form:
\begin{equation}\label{11}
\textbf{A}_{a/b}= C^c_{ba}\textbf{A}_c.
\end{equation}
Throughout the text the following notations are used:
$$
\xi_a^\alpha F_{,\alpha}=F_{/a}, \quad \zeta_a^\alpha F_{,\alpha}=F_{|a}.
$$
Next notation
\begin{equation}\label{12}
\mathbb{A}_a =\zeta_a^\alpha A_\alpha
\end{equation}
is used for the nonholonomic components of the of the electromagnetic field vector potential in the canonical frame. Throughout the text, functions denoted by lowercase Greek letters depend only on the variable $u^0$.

\quad

\noindent
The {\bf LEMMA} takes place.

\quad

\noindent

The nonholonomic components\eqref{12} of the electromagnetic field vector potential in the canonical frame depend only on $u^0$:
\begin{equation}\label{13}
\mathbb{A}_{a}=\alpha_{a}(u^0).
\end{equation}
\noindent
{\bf Proof.}

\quad

\noindent
The right side of the system of equations \eqref{10} can be represented in the following form:
\begin{equation}\label{14}
(\xi_a^\alpha A_\alpha)_{/b}=(\xi_a^\alpha \zeta_\alpha^c \mathbb{A}_c)_{/b} =
\xi_a^\alpha \zeta_\alpha^c \mathbb{A}_{c/b}.
+ \xi_{a/b}^\alpha \zeta_{\alpha}^c \mathbb{A}_c+\xi_{a}^\alpha \zeta_{\alpha/b}^c \mathbb{A}_c
\end{equation}
From \eqref{6} it can obtain the equations:
$$
\zeta_{\beta,\alpha}^a = -\zeta_{\gamma}^a \xi^b_\alpha \xi^\gamma_{b,\beta}.
$$
Using this, one can reduce the expression \eqref{14} to the form:
\begin{equation}\label{15}
\textbf{A}_{a/b}=\mathbb{A}_{c/a}\xi^\alpha_b\zeta^c_\alpha +(\xi^\alpha_a\xi^\gamma_{b,\alpha}-\xi^\alpha_b\xi^\gamma_{a,\alpha})\zeta^c_\gamma\mathbb{A}_c = \mathbb{A}_{c/a}\xi^\alpha_b\zeta^c_\alpha + C^c_{ab}\textbf{A}_c.
\end{equation}
Thus, from \eqref{10}, \eqref{14},  \eqref{15} it follows:
\begin{equation}\label{16}
\mathbb{A}_{a/b}=0 \Rightarrow \mathbb{A}_{a}=\alpha_{a}(u^0).
\end{equation}

\noindent
The {\bf LEMMA} is proved.

\quad

\noindent

As it is known (see \cite{78}), the non-holonomic components of the Ricci tensor in the canonical frame depend only on the functions $\eta_{\alpha\beta}$ and the structural constants. A similar situation occurs for the energy-momentum tensor of the electromagnetic field, invariant with respect to the group\quad $G_3(N)$.\quad Let us show that in the non null homogeneous Petrov space of type $V_4(N)$ the Maxwell vacuum equations:
\begin{equation}\label{17}
\frac{1}{G}(G F^{ij})_{,j}=0
\end{equation}
are reduced to a system of algebraic equations on the function $\eta_{\alpha\beta}, $ which includes the functions $\alpha_a$ also the structure constants.
Let's denote the components $F_{ab}$ of the electromagnetic field tensor in the canonical frame $\zeta_a^\alpha$ as follows:
$$
F_{ab}=\zeta_a^\alpha\zeta_b^\beta F_{\alpha\beta}, \quad F_{0a}=\zeta_a^\alpha F_{0\alpha}=\dot{\alpha}_a \quad F^{ab}=\eta^{aa_1}\eta^{bb_1}F_{a_1b_1}
$$
From condition \eqref{11} it follows:
$$
F_{ab} = C_{ba}^c\alpha_c \Rightarrow F^{ab}=-{\eta^2}((\eta^{a2}\eta^{b3} -\eta^{a3}\eta^{b2})C_{23}^c\alpha_c +(\eta^{a3}\eta{b1} -\eta^{a1}\eta^{b3})C_{31}^c\alpha_c +(\eta^{a1}\eta{b2} -\eta^{a2}\eta^{b1})C_{12}^c\alpha_c.
$$
Denoting:
$$
f^1=C_{23}^c\alpha_c = C^{1c}\alpha_c, \quad f^2=C_{31}^c\alpha_c =C^{2c}\alpha_c, \quad f^3=C_{12}^c\alpha_c =C^{3c}\alpha_c, \quad f_1=F^{23}, \quad f_2=F^{31}, \quad f_3=F^{12},
$$
one can reduce these relations to the form:
$$
f_a =\frac{\varepsilon}{\eta^2}\eta_{ab}f^b.
$$
Let us consider the following equations from the system of Maxwell equations:
\begin{equation}\label{18}
\frac{\zeta_\alpha^a}{G}(G F^{\alpha i})_{,i}=\frac{\zeta_\alpha^a}{\zeta}(\zeta \zeta^\alpha_{a_1} \zeta^\beta_b F^{a_1b} )_{,\beta} + \frac{1}{\eta}(\eta F^{a 0})_{,0}=F^{ab}C^c_{cb}+\frac{1}{2} F^{cb}C^a_{bc} + \frac{1}{\eta} (\eta\eta^{ab}\dot{\alpha}_b)_{,0}.
\end{equation}
Here we use the expression for the first term on the right-hand side.
$$
\frac{\zeta_\alpha^a}{\zeta}(\zeta \zeta^\alpha_{a_1} \zeta^\beta_b F^{a_1b} )_{,\beta}= F^{ab}(\zeta^\alpha_{b,\alpha}-\zeta^\alpha_{c|b}\zeta^c_\alpha)+F^{cb}\zeta^\alpha_{c,|b}\zeta^a_\alpha = F^{ab}C^c_{cb}+\frac{1}{2} F^{cb}C^a_{bc}
$$
In order to reduce Maxwell equations to a system of ordinary differential equations of the first order, we introduce new independent functions:
$$
\beta^a=\eta\eta^{ab}\dot{\alpha}_b.
$$
In this case, the system of Maxwell equations must be supplemented with the system of equations:
\begin{equation}\label{19}
\dot{\alpha}_b=\frac{1}{\eta}\eta_{ab}\beta^a.
\quad
\end{equation}
The system of equations \eqref{18} will take the form:
\begin{equation}\label{20}
\dot{\beta^a}+\frac{1}{\eta}f^b(\eta_{1b}(C^{1a}- C^a_{a3}\delta_2^a)+\eta_{2b}(C^{2a}+ C^a_{a3}\delta_1^a)+\eta_{3b}C^{3a})=0.
\end{equation}
These equations \eqref{20} should be supplemented by the equation:
\begin{equation}\label{21}
C^a_{ab}\beta^b =0
\end{equation}
(it follows from the equations \quad $\frac{1}{g}(g F^{0\alpha})_{,\alpha}=0 $). \quad Denoting
$$
n_{ab}= \frac{1}{\eta}\eta_{ab}, \quad Q= C^a_{a3},
$$
one can represent the system of Maxwell equations as the following system of algebraic (with respect to the functions $n_{\alpha\beta}$) equations:
\begin{equation}\label{22}
n_{ab}\beta^b=\dot{\alpha}_a,
\end{equation}
\begin{equation}\label{23}
f^b(n_{1b}(C^{1a}- Q\delta_2^a)+n_{2b}(C^{2a}+Q\delta_1^a)+n_{3b}C^{3a})=-\dot{\beta^a},
\end{equation}
\begin{equation}\label{24}
Q\beta^b =0.
\end{equation}
For convenience, we represent in the table the values of all quantities included in the equations \eqref{22} - \eqref{24}.
Depending on the number $N$ of the group $G_3(N)$, they have the following form:

\begin{equation}\label{25}
\begin{tabular}{|l|l|l|l|l|l|l|l|l|l|}
\hline
$N$ & $ I $ & $II$ & $ III $ & $ IV $ & $ V $ & $ VI $ & $ VII $ & $VIII $ & $IX $ \\
\hline
$C^{11}$& $ 0 $ & $ 1 $ & $ 0 $ & $ 1 $ & $ 0 $ & $ 0 $ & $ -1 $ & $ 0 $ & $ 1 $ \\
\hline
$C^{12}$& $ 0 $ & $ 0 $ & $ 0 $ & $ 1 $ & $ 1 $ & $ q $ & $ q $ & $ 0 $ & $ 0 $ \\
\hline
$C^{21}$& $ 0 $ & $ 0 $ & $ -1 $ & $ -1 $ & $ -1 $ & $ -1 $ & $ 0 $ & $ 0 $ & $ 0 $ \\
\hline
$C^{22}$& $ 0 $ & $ 0 $ & $ 0 $ & $ 0 $ & $ 0 $ & $ 0 $ & $ -1 $ & $ -1 $ & $ 1 $ \\
\hline
$C^{33}$& $ 0 $ & $ 0 $ & $ 0 $ & $ 0 $ & $ 0 $ & $ 0 $ & $ 0 $ & $ 0 $ & $ 1 $ \\
\hline
$C^{21}+Q$& $ 0 $ & $ 0 $ & $ 0 $ & $ 1 $ & $ 1 $ & $ q $ & $ q $ & $ 0 $ & $ 0 $ \\

\hline
$C^{12}-Q $ & $ 0 $ & $ 0 $ & $ -1 $ & $ -1 $ & $ -1 $ & $ -1 $ & $ 0 $ & $ 0 $ & $ 0 $ \\

\hline
$C^{13}$& $ 0 $ & $ 0 $ & $ 0 $ & $ 0 $ & $ 0 $ & $ 0 $ & $ 0 $ & $ 1 $ & $ 0 $ \\

\hline
$C^{31}$& $ 0 $ & $ 0 $ & $ 0 $ & $ 0 $ & $ 0 $ & $ 0 $ & $ 0 $ & $ 1 $ & $ 0 $ \\

\hline
$f^1$& $ 0 $ & $\alpha_1$ & $0$ & $\alpha_1 +\alpha_2$ & $ \alpha_2 $ & $ q\alpha_2 $ & $ q\alpha_2 -\alpha_1 $ & $ \alpha_3 $ & $ \alpha_1 $ \\
\hline
$f^2$& $ 0 $ & $ 0 $ & $-\alpha_1$ & $-\alpha_1$ & $- \alpha_1 $ & $ -\alpha_1 $ & $ -\alpha_2
$ & $ -\alpha_2 $ & $ \alpha_2 $ \\
\hline
$f^3$& $ 0 $ & $ 0 $ & $0$ & $0$ & $ 0 $ & $ 0 $ & $ 0
$ & $ \alpha_1 $ & $ \alpha_3 $ \\
\hline
$Q$ & $ 0 $ & $ 0 $ & $ 1 $ & $ 2 $ & $ 2 $ & $ q+1 $ & $ q
$ & $ 0 $ & $ 0 $ \\
\hline
$\Delta(N)$& $ 0 $ & $0$& $ 0 $ & $ 1 $ & $ 1 $ & $ q $ & $ 1 $ & $0$ & $1$ \\
\hline
\end{tabular}
\end{equation}
Here \quad $\Delta(N)=C^{11}C^{22}+(Q+C^{21})(Q-C^{12}).$

\quad

\noindent
The set of non-null homogeneous Petrov spaces can be divided into three subsets:

\quad

\noindent
${\bf I.} $ Spaces with solvable groups of motions \quad $G_3(I)-G_3(III) \quad (\beta^3 = const);$

\quad

\noindent
${\bf II.} $ Spaces with solvable groups of motions \quad $G_3(IV)-G_3(VII) \quad (\beta^3 = 0, \quad Q\Delta(N) \ne 0);$

\quad

\noindent
${\bf III.} $ Spaces with non-solvable groups of motions \quad $G_3(VIII)-G_3(IX) \quad (Q\Delta(N) =0).$

\quad

\noindent
In this paper, a unified approach to implementing classification for all types of nonzero homogeneous Petrov spaces is used. The classification is based on the use of solutions of the system of equations \eqref{22} at the initial stage. The set of these solutions (having a common form for all spaces) can be divided into three subsets. Let us represent all these subsets.

\quad

\noindent
Option {\bf A}\quad $\beta^3 \ne 0$. \quad The option is not realized for the groups \quad $G_3(III)- G_3(VII)$,\quad since for these groups \quad $ Q \ne 0 \Rightarrow \beta^3=0$. \quad It will also not be used as the initial classification step for the group \quad $G_3(VIII),$ \quad since in this case it is appropriate to start the classification with the option \quad $\beta^2 \ne 0.$  \quad
The solution of the system of equations \eqref{22} has the form:
\begin{equation}\label{26}
n_{3p}=\frac{1}{\beta^3}(\dot{\alpha}_p - n_{pq}\beta^q), \quad n_{33}=\frac{1}{{\beta^3}^2}(\dot{\alpha}_3\beta^3 +\dot{\alpha}_p\beta^p - n_{pq}\beta^q \beta^q) \quad (p,q = 1,2).
\end{equation}
Note that all functions on the right-hand sides of the equations \eqref{26} at this stage are considered as independent functions of the variable $u^0$. After substituting \eqref{26} into the system of equations \eqref{23}, some of them will remain independent. The other functions will be expressed through these independent functions, as well as through independent functions $\gamma(u^0), \omega(u^0$), etc., which be introduced to reduce the order of the system of differential equations \eqref{23}. This remark is also true for all other types of spaces.

\quad

\noindent
Option {\bf B} \quad $\beta^2 \ne 0$.\quad For solvable groups $G_3(IV)$ - $G_3(VII)$ one has the additional condition \quad $\beta^3=0$.

Then the solution of the system of equations \eqref{22} has the form:
\begin{equation}\label{27}
n_{12}=\frac{1}{\beta^2}(\dot{\alpha}_1 - n_{11}\beta^1), \quad n_{22}=\frac{1}{{\beta^2}^2}(\dot{\alpha}_2\beta^2 -\dot{\alpha}_1\beta^1 + n_{11}{\beta^1}^2),\quad n_{23}=\frac{1}{\beta^2}(\dot{\alpha}_3 - n_{13}\beta^1).
\end{equation}

\quad

\noindent
Option {\bf C}  \quad $\beta^2 = \beta^3=0$. \quad For the groups $G_3(V)$, $G_3(VI)$, $G_3(VIII)$, there exist admissible transformations of canonical vectors that preserve the form of the structure equations. These transformations have the form:

\quad

\noindent
--for the group $G_3(V): \quad \tilde{\zeta_1^\alpha}= \zeta_2^\alpha, \quad \tilde{\zeta_2^\alpha}= \zeta_1^\alpha $;

\quad

\noindent
--for the group $G_3(VI): \quad \tilde{\zeta_1^\alpha}= \zeta_2^\alpha, \quad \tilde{\zeta_2^\alpha}= \zeta_1^\alpha,\quad \tilde{\zeta_3^\alpha}= q \zeta_3^\alpha, \quad \tilde{q}= \frac{1}{q} $;

\quad

\noindent
--for the group $G_3(VIII): \quad \tilde{\zeta_1^\alpha}= \zeta_3^\alpha, \quad \tilde{\zeta_3^\alpha}= \zeta_1^\alpha $.

\quad

\noindent
Therefore, for these groups, the option {\bf C} is a special case of the option {\bf B}, and it does not  considered separately. For the remaining groups, the solution of the system of equations \eqref{22} has the form:

\begin{equation}\label{28}
n_{1a}=\frac{1}{\beta^1}\dot{\alpha}_a.
\end{equation}

\noindent
Option {\bf D}  \quad $\beta^a = 0.$ \quad From the equations \eqref{22} for all solvable groups groups $G_3(N)$ it follows:
\quad
$\alpha_a=a_a = const. $  \quad
For non-solvable groups, the electromagnetic field is absent.

\quad

\noindent
Holonomic components of the metric tensor and vector potential of the electromagnetic field are given by the formulas \eqref{7}, \eqref{12}:
\begin{equation}\label{29}
g_{\alpha\beta}= {\zeta_\alpha}^{a}{\zeta_\beta}^{b} \eta^{ab}(u^0),\quad
 A_\alpha=\zeta^a_\alpha\mathbb{A}_a = \zeta^a_\alpha \alpha_a.
\end{equation}
Since for all homogeneous Petrov spaces the pairs of vector fields $\xi^\alpha_a $ and $\zeta^\alpha_a $ have already been found (see \cite{12}), the solutions of the vacuum Maxwell equations in the holonomic coordinate system are determined by the functions $n_{ab},\alpha_a$. The types of these functions are determined from the solution of the system of equations \eqref{22}-\eqref{24} and are given below for each Petrov space.

\quad

\noindent
The following notations are used below:

\quad

\noindent
$$\varsigma,\varsigma_a, \xi, \xi_a = \pm 1, \quad a, a_a, b, b_a, c, c_a = const. $$

\quad

\noindent

\section{SOLUTIONS OF MAXWELL EQUATIONS FOR GROUPS \quad ${\bf G_3(N<IV)}$}

\quad

\noindent
\textbf{4.1} \quad \textbf{ GROUP}  $\textbf{G}_{\textbf{3}}$\textbf{(I)}

\quad

\noindent
 $G_3(I)$ is an Abelian group, and all structure constants are zero $\Rightarrow \xi^\alpha_a = \zeta^\alpha_a = \delta^\alpha_a$.  From Maxwell equations it follows:
$$
\beta^a = const.
$$
Vector $\beta^a $ can be diagonalized by admissible coordinate transformations:
$$
\beta^a = \frac{1}{c}\delta^a_3.
$$
The solution can be written as:
$$
\eta_{ab}=\frac{\varepsilon n_{ab}}{\det\left\|n_{a_1b_1}\right\|}, \quad n_{a3}= c\dot{\alpha}_a.
$$
$\alpha_a$ and other components $n_{ab}$ are arbitrary functions.
Note that in the case of the Abelian group, holonomic and nonholonomic coordinate systems coincide.

\quad

\noindent
\textbf{4.2} \quad \textbf{GROUP}  $\textbf{G}_{\textbf{3}}$\textbf{(II)}

\quad

\noindent
The Killing vector fields  $\xi^\alpha_a $ and the vector fields  of the canonical frame   $\zeta^\alpha_a $  (as it has been already noted, these vectors can be swapped) have the form :
\begin{equation} \label{30}
\xi_a^\alpha = \delta_a^1 \delta_1^\alpha + \delta_a^2 \delta_2^\alpha +\delta_a^3 (\delta_3^\alpha +u^2\delta_1^\alpha).
\end{equation}
$$
\zeta_a^{\alpha } =\delta _{1}^{\alpha } \left(\delta _{a}^{1} +u^{3} \delta _{a}^{3} \right)+\delta _{2}^{\alpha } \delta _{a}^{2} +\delta _{3}^{\alpha } \delta _{a}^{3}.
$$

\quad

\noindent
The equations \eqref{23} are reduced to the following:
\begin{equation}\label{31}
\dot{\beta}^a=-\delta^a_1\alpha_1 n_{11} \Rightarrow \dot{\beta}^1=-\alpha_1 n_{11}, \quad \beta^2 = c_2, \quad \beta^3=c_3 \quad (c_a=const).
\end{equation}
Let's consider all possible options.

\quad

\noindent
Option ${\bf A.}\quad \beta^3=1.$ \quad Using admissible transformations of vectors $\zeta_p^\alpha \quad (p, q = 2,3), $ one of the parameters $c_p$ (for example, $c_2=\beta^2$) can be set to zero. Then the equations \eqref{22} takes the form:
$$
n_{13}=\dot{\alpha}_1-n_{11}\beta^1, \quad n_{23}=\dot{\alpha}_2-n_{12}\beta^1, \quad n_{33}=\dot{\alpha}_3-\dot{\alpha}_1\beta^1 + n_{11}{\beta^1}^2.
$$
The equations \eqref{31}, has next solutions:

\quad

\noindent

Variant ${\bf A_1}\quad \alpha_1 \ne 0$,
  $$n_{11}=\frac{1}{\alpha_1}\dot{\beta}^1, \quad n_{13}=\frac{\dot{\alpha}_1\alpha_1+\beta^1\dot{\beta}_1}{\alpha_1}, \quad n_{23}=\dot{\alpha}_2-n_{12}\beta^1,\quad
n_{33}=\dot{\alpha}_3-\beta^1\frac{\dot{\alpha}_1\alpha_1+\beta^1\dot{\beta}_1}{\alpha_1}.
$$

\quad

\noindent

Variant ${\bf A_2} \quad\alpha^1=0,\quad \beta^1 =c,$
$$
n_{13}=-c n_{11}, \quad n_{23}=\dot{\alpha}_2 -c n_{12}, \quad n_{33}=\dot{\alpha}_3 +c^2 n_{11},
$$

$n_{11}$ is an arbitrary function.

\quad

\noindent
Option ${\bf B} \quad \beta^3=0, \quad \beta^2 = 1.$ \quad It is equivalent to the previous option.

\quad

\noindent
Option ${\bf C}\quad \beta^3=\beta^2 = 0,\quad \beta^1 \ne 0, \quad n_{1a}=\frac{1}{\beta^1}\dot{\alpha_a}.$ \quad Solution of the equation \eqref{31} have the form:
$$
\alpha_1=a \sin\omega, \quad \beta^1= a\cos\omega, \quad n_{11}=\dot{\omega}, \quad n_{12}=\frac{\dot{\alpha_2}}{a \cos\omega},\quad n_{13}=\frac{\dot{\alpha_3}}{a \cos\omega}.
$$

\quad

\noindent
\textbf{4.3} \quad \textbf{GROUP}  $\textbf{G}_{\textbf{3}}$\textbf{(III)}

\quad

\noindent
The Killing vector fields $\xi^\alpha_a $ and the canonical frame $\zeta^\alpha_a $ can be represented in the form:
\begin{equation} \label{32}
\xi_a^\alpha = \delta_a^1 \delta_1^\alpha + \delta_a^2 \delta_2^\alpha +\delta_a^3 (\delta_3^\alpha +u^2\delta_2^\alpha),
\end{equation}
$$
\zeta_{\left(a\right)}^{\alpha } =\delta _{1}^{\alpha } \delta _{a}^{1} +\delta _{2}^{\alpha } \delta _{a}^{2} \exp \left(-u_{3} \right)+\delta _{3}^{\alpha } \delta _{a}^{3}.
$$

\noindent
Equations \eqref{23} have the form:
\begin{equation}\label{33}
\dot{\beta}^1=0 \Rightarrow \beta^1 = c =const, \quad \dot{\beta}^2=-\alpha_1 n_{12}.
\end{equation}
Since in this case from equation \eqref{33} it follows \quad $\beta^3=0$, \quad option ${\bf A}$  do not be realized. Therefore, only options ${\bf B, C, D}$ should be considered.

\quad

\noindent
Option ${\bf B} \quad \beta^2 \ne 0 , \quad \beta^1=c.$ \quad

\quad

\noindent
Equations \eqref{33} need to be supplemented with equations \eqref{22},having in this case the form:
\begin{equation}\label{34}
n_{12} = \frac{1}{\beta^2}(\dot{\alpha}_1-c n_{11}), \quad n_{22}= \frac{1}{{\beta^2}^2}(\dot{\alpha}_2\beta^2-\dot{\alpha}_1 c + c^2 n_{11}), \quad n_{23}=\frac{1}{\beta^2}(\dot{\alpha}_3 - n_{13}c).
\end{equation}
From the relations \eqref{33}, \eqref{34} it follows
$$
({\alpha_1}^2 + {\beta^2}^2)_{,0}=2c\alpha_1 n_{11}.
$$
Using this relation, one finds all the corresponding solutions of Maxwell equations.

\quad

\noindent

Variant ${\bf B_1}$\quad $ \alpha_1 \ne 0 \Rightarrow n_{21} = -\frac{\dot{\beta^2}}{\alpha_1}, \quad cn_{11}=\frac{\dot{\alpha_1}\alpha_1+\beta^2\dot{\beta^2}}{\alpha_1};$

\quad

\noindent
${1}\quad
 \beta^1 = c = 0,\quad \alpha_1=a \sin\omega, \quad \beta^2= a \cos\omega, \quad n_{12}=\dot{\omega}, \quad n_{11}$ \quad is an arbitrary function.

\quad

\noindent
$2\quad \beta^1 =c, \quad \beta^1 = c \Rightarrow \beta^2 =1. \quad n_{11}=\frac{\dot{\alpha_1}\alpha_1+\beta^2\dot{\beta^2}}{c\alpha_1}\quad  n_{12}=\dot{\omega}, \quad n_{12}$ \quad is an arbitrary function.

\quad

\noindent

Variant  ${\bf B_2 } \quad \alpha_1 = 0\Rightarrow\beta^2 =1,
\quad n_{12}=- c n_{11}.\quad \quad n_{11}$ \quad is an arbitrary function.

\quad

\noindent
Option {$\bf{C}$}\quad $\beta^2=0$. \quad From equations \eqref{22}, \eqref{28} it follows: \quad $\beta^1=1, \quad \alpha_1 n_{12}=0, \quad n_{1a}=\dot{\alpha_a}$. The solution has the form:

\quad

\noindent
{$1$}\quad $ \alpha_2 = a, \quad n_{12}=0, $

\quad

\noindent
$2\quad
\alpha_1 =0, \quad n_{11}=0. \quad n_{12}$ \quad is an arbitrary function.

\quad

\noindent
Option $\bf {D}$\quad $\alpha_a=const, \quad \beta^a=0$. \quad From equation \eqref{35} there follow two solutions of equations \eqref{23}:

\quad

\noindent
 $1$\quad $\alpha_1 =0,$

\quad

\noindent
$2$\quad $n_{12}=0.$

\quad

\noindent
\section{\textbf{SOLUTIONS OF MAXWELL EQUATIONS FOR THE GROUPS} $\textbf{G}_\textbf{3}\textbf{(N)}$\quad \textbf{$(III < N < VIII)$}}

\quad

\noindent
From the system of equations \eqref{23} it follows:
\begin{equation}\label{35}
n_{11}f^1 + n_{12}f^2 =\frac{\dot{\beta^2}(Q+C^{21})-\dot{\beta^1}C^{22}}{\Delta(N)}, \quad
n_{12}f^1 + n_{22}f^2 = \frac{\dot{\beta^1}(C^{12}-Q)-\dot{\beta^2}C^{11}}{\Delta(N)}.
\end{equation}
Since Option $\textbf{A}$ does not be realized ($\beta^3 =0$), only Option $\textbf{B, C, D}$ should be considered.

\quad

\noindent
Option {\bf B.} \quad $\beta^2 \ne 0, \quad \beta^3=0$,\quad

\quad

\noindent
From equations \eqref{22} we find additional conditions on functions $n_{2a}:$
$$
n_{22}=\frac{1}{{\beta^2}^2}(\dot{\alpha}_2\beta^2 -\dot{\alpha}_1\beta^1 + n_{11}{\beta^1}^2), \quad n_{12}=\frac{1}{\beta^2}(\dot{\alpha}_1 - n_{11}\beta^1), \quad n_{23}=\frac{1}{\beta^2}(\dot{\alpha}_3 - n_{12}\beta^1).
$$
Using these conditions in the system of equations \eqref{35}, we reduce this system to the form:
\begin{equation}\label{36}
n_{11}(\beta^2 f^1 -\beta^1f^2) =\frac{\beta^2}{\Delta}(C^{22}\dot{\beta}^1 -(C^{21} + Q)\dot{\beta}^2) - f^2 \dot{\alpha}_1, \quad
\end{equation}
\begin{equation}\label{37}
\Delta(N)(\dot{\alpha}_1f^1 + \dot{\alpha}_2f^2) = (Q - C^{12})\dot{\beta}^1\beta^2 + C^{11}\dot{\beta}^2\beta^2 +C^{22}\dot{\beta}^1\beta^1 -(C^{21} + Q)\dot{\beta}^2 \beta^1.
\end{equation}
Thus, the solutions of Maxwell equations for this Opyion ${\bf B}$ fall into two types

\quad

\noindent

Variant {\bf$ B_1. \quad (\beta^2 f^1 -\beta^1f^2) \ne 0 \Rightarrow$}
\begin{equation}\label{38}
n_{11} =\frac{\beta_2 }{\Delta(N)(\beta^2 f^1 -\beta^1f^2)}(C^{22}\dot{\beta}^1 -(C^{21} + Q)\dot{\beta}^2) -f^2 \dot{\alpha}_1). \quad
\end{equation}
The functions $\alpha_p, \beta^p$ are related by the condition \eqref{37}.

\quad

\noindent
Variant {\bf$ B_2 \quad (\beta^2 f^1 -\beta^1f^2) = 0 $},

\quad

\noindent
$n_{11}$ is an arbitrary function. In addition to the condition \eqref{37}, two conditions more are imposed on the functions $\alpha_p, \beta^p$:
\begin{equation}\label{39}
\beta^2 f^1 =\beta^1f^2, \quad \beta^2(C^{22}\dot{\beta}^1 -(C^{21} + Q)\dot{\beta}^2) =f^2 \dot{\alpha}_1\Delta(N).
\end{equation}

\quad

\noindent
Option {\bf {C} $ \quad \beta^2 =\beta^3 =0.$}

\quad

\noindent
From the conditions \eqref{23} it follows \quad $n_{11} = \frac{\dot{\alpha}_1}{\beta^1}, \quad n_{12} = \frac{\dot{\alpha}_2}{\beta^1},\quad n_{13} = \frac{\dot{\alpha}_3}{\beta^1}$ \quad Substitute these expressions into \eqref{35}. As a result, we obtain:
\begin{equation}\label{40}
n_{22}f^2 =\frac{1}{\Delta}(Q - C^{12})\dot{\beta}^1 -f^1 \dot{\alpha}_2,
\end{equation}
\begin{equation}\label{41}
\frac{1}{\Delta} C^{22}\dot{\beta}^1= f^1 \dot{\alpha}_1 + f^2 \dot{\alpha}_2.
\end{equation}
Thus, for this, two variants of solutions are possible.

\quad

\noindent

Variant {\bf$ C_1$} \quad $f^2 \ne 0$.
\begin{equation}\label{42}
n_{22} =\frac{1}{f^2}(\frac{1}{\Delta}(Q - C^{12})\dot{\beta}^1 -f^1 \dot{\alpha}_2),
\end{equation}
The functions $\alpha_p, \beta^1$ are related by the condition \eqref{41}.

\quad

\noindent

Variant {\bf$ C_2$} \quad $f^2=0 \quad \Rightarrow n_{22} - $ arbitrary function of $u^0$.

\quad

\noindent
In addition to the condition \eqref{41}, another condition is imposed on the functions $\alpha_p, \beta^1$, which follows from the equation \eqref{40}:
\begin{equation}\label{43}
\frac{1}{\Delta}(Q - C^{12})\dot{\beta}^1 = f^1 \dot{\alpha}_2.
\end{equation}

\quad

\noindent
Option {\bf D} $ \quad \beta^a =0\Rightarrow \alpha_a=const.$

\quad

\noindent In this case, the system \eqref{23} can be represented as:
\begin{equation}\label{44}
n_{11}f^1 + n_{12}f^2 =0,
\end{equation}
$$
n_{12}f^1 + n_{22}f^2 = 0.
$$
System \eqref{44} must have nonzero solutions, otherwise the electromagnetic field is absent. Therefore
$$
\det\left\|n_{pq}\right \|  \Rightarrow n_{pq} = {\gamma_p}\gamma_q \Rightarrow \gamma_p f^p =0 \quad (\gamma_p = {\gamma_p}(u^0) ).
$$

\quad

\noindent
\textbf{5.1}. \quad\textbf{ GROUP} $\textbf{G}_\textbf{3}\textbf{(IV)}$

\quad

\noindent
Let us present the Killing vector fields:
\begin{equation} \label{45}
\xi _{a}^{\alpha } =\delta _{1}^\alpha \delta^1_a\exp(-u^3) + \delta _{2}^{\alpha }\delta_a^2 +\delta _{a}^{3}( \delta_3^\alpha +u^{2}(\delta_2^\alpha + \delta_{1}^{\alpha}\exp(-u^3)).
\end{equation}
and the vector fields of the canonical frame:
$$
\zeta_{a}^{\alpha } =\delta _{1}^\alpha (\delta^1_a + u^2\delta_a^3)\exp( -u^3) + \delta _{2}^\alpha (\delta^2_a + u^2\delta_a^3)
+\delta _{a}^{3}\delta_3^\alpha .
$$

\quad

\noindent
The system of equations \eqref{36} has the form:
\begin{equation}\label{46}
n_{12}(\alpha_1 + \alpha_2) - n_{22}\alpha_1 =-(\dot{\beta}^1 +\dot{\beta}^2), \quad
\end{equation}
$$
n_{11}(\alpha_1 + \alpha_2) - n_{12}\alpha_1 =\dot{\beta}^2.
$$
Consider the

\quad

\noindent
Option ${\bf B}$.

\quad

\noindent
Components $n_{p2} \quad (p, q =1,2)$ have the form (see \eqref{22}):
$$
n_{12}=\frac{1}{\beta^2}(\dot{\alpha}_1 - n_{11}\beta^1), \quad n_{22}=\frac{1}{{\beta^2}^2}(\dot{\alpha}_2\beta^2 -\dot{\alpha}_1\beta^1 + n_{11}{\beta^1}^2).
$$
Using the conditions \eqref{22}, \eqref{37}, one can represent equations \eqref{46} in the form:
\begin{equation}\label{47}
n_{11}(\beta^2 (\alpha_1 + \alpha_2) + \beta^1\alpha_1) =\beta^2\dot{\beta}^2 + \alpha_1\dot{\alpha}_1, \quad
\end{equation}
\begin{equation}\label{48}
\dot{\alpha}_1(\alpha_1 + \alpha_2) - \dot{\alpha}_2\alpha_1 + (\dot{\beta}^1 + \dot{\beta}^2)\beta^2 -\dot{\beta}^2 \beta^1 =0.
\end{equation}

\quad

\noindent
Let's consider all possible variants.

\quad

\noindent

Variant ${\bf B_1} \quad \beta^2 (\alpha_1 + \alpha_2) + \beta^1\alpha_1 \ne 0.$ \quad In this case:
\begin{equation}\label{49}
n_{11} = \frac{1}{(\beta^2 (\alpha_1 + \alpha_2) + \beta^1\alpha_1)}(\beta^2\dot{\beta}^2 + \alpha_1\dot{\alpha}_1). \quad
\end{equation}
If the function $\alpha_1 =0,$ one can reduced equation \eqref{47} to the form:
\begin{equation}\label{50}
(\ln\beta^2 +\frac{\beta_1}{\beta^2} )_{,0} =0.
\end{equation}
Hence the first solution is:

\quad

\noindent
1 \quad $\alpha_1 =0, \quad \beta_1 = \beta_2 (c -\ln \beta_2). $

\quad

\noindent Consider the case when \quad $\alpha_1 \ne 0.$ \quad
Let us introduce new independent functions as follows:
$$
\alpha_1 =\exp\sigma \cos\omega,\quad \beta^2 =\exp\sigma\sin\omega, \quad \alpha_2=\gamma_2\cos\omega, \quad \beta^1 = \gamma_1 \sin\omega.
$$
The equation \eqref{48} will take the form:
\begin{equation}\label{51}
\cos^2(\omega)(\dot{\gamma_1} + \dot{\gamma_2}) = \dot{\sigma} + \dot{\gamma_1} \Rightarrow \sin^2(\omega)(\dot{\gamma_1} + \dot{\gamma_2}) = \dot{\gamma_2} -\dot{\sigma}.
\end{equation}
The equation \eqref{51} has two independent solutions:

\quad

\noindent
2 \quad $
\alpha_1 =\exp\sigma \cos\omega,\quad \beta^2 =\exp\sigma\sin\omega, \quad \alpha_2=\sigma \exp\sigma \cos\omega, \quad \beta^1 = -\sigma \exp\sigma\sin\omega.
$

\quad

\noindent
$ 3 \quad \cos(\omega)= \sqrt{\frac{\dot{\sigma}+\dot{\gamma_1}}{\dot{\gamma_1}+\dot{\gamma_2}}}, \quad \sin(\omega) =\sqrt{\frac{\dot{\gamma_2}-\dot{\sigma}}{\dot{\gamma_2}+\dot{\gamma_1}}} \Rightarrow $
$$\alpha_1= \exp{\sigma} \sqrt{\frac{\dot{\sigma}+\dot{\gamma_1}}{\dot{\gamma_1}+\dot{\gamma_2}}}, \quad \beta^2 =\exp{\sigma} \sqrt{\frac{\dot{\gamma_2}-\dot{\sigma}}{\dot{\gamma_2}+\dot{\gamma_1}}}, \quad,
$$
$$\alpha_2 =\gamma_2\exp{\sigma} \sqrt{\frac{\dot{\sigma}+\dot{\gamma_1}}{\dot{\gamma_1}+\dot{\gamma_2}}}, \quad \beta^1 =\gamma_1\exp{\sigma} \sqrt{\frac{\dot{\gamma_2}-\dot{\sigma}}{\dot{\gamma_2}+\dot{\gamma_1}}} $$

\quad

\noindent

Variant ${\bf B_2} \quad n_{11}$  is an arbitrary function, \quad $\beta^2\dot{\beta}^2 - \alpha_1\dot{\alpha}_1=0 \Rightarrow \alpha_1 =p \cos\omega, \quad \beta^2 =p\sin\omega$. \quad The functions \quad $\alpha_2, \quad\beta^1 $ \quad obey the equations:
\begin{equation}\label{52}
{\alpha_1}^2(\frac{\dot{\alpha}_2}{\alpha_1})_{,0} = {\beta^2}^2(\frac{\dot{\beta}^2}{\beta^2})_{,0}, \quad \beta^2 (\alpha_1 + \alpha_2) + \beta^1\alpha_1 = 0. \quad
\end{equation}
\quad

\noindent
The solution of the system of equations \eqref{52} can be represented as following:

\quad

\noindent
4. \quad $
\alpha_1 =(c_1 + c_2)\cos\omega,\quad \beta^2 =(c_1 + c_2)\sin\omega, \quad \alpha_2=-c_2\cos\omega, \quad \beta^1 = -c_1\sin\omega.$

%
%
%
%
%
%

\quad

\noindent
Let's consider the

\quad

\noindent
Option {\bf C} \quad $\beta^1 \ne 0, \quad \beta^2=0.\quad n_{1a}=\frac{\alpha_a}{\beta^1}. $ \quad Maxwell  equations takes the form:
\begin{equation}\label{53}
\dot{\alpha}_1(\alpha_1 + \alpha_2) - \dot{\alpha}_2\alpha_1 = 0, \quad n_{22}\beta^1\alpha_1=(\dot{\alpha}_2\alpha_2+\dot{\beta}^1\beta^1) +\dot{ \alpha_2}\alpha_1.
\end{equation}
There are two solutions of the equations \eqref{53}:

\quad

\noindent
${5}$ \quad $\alpha_1 = 0,\quad n_{11} =0, \quad n_{12} =\dot{\omega}.\quad
\alpha_2=a\cos\omega, \quad \beta^1 = a \sin\omega.
$ \quad $n_{22}$ is an arbitrary function.

\quad

\noindent
${6} \quad n_{22}=\frac{1}{\beta^1\alpha_1}(\dot{\alpha}_2\alpha_2+\dot{\beta}^1\beta^1 +\dot{ \alpha_2}\alpha_1).\quad \alpha_2 = \alpha_1(c+\ln{\alpha_1}), \quad n_{11}$ is an arbitrary function.

\quad

\noindent
Let's consider the:

\quad

\noindent
 Option {\bf D} \quad $\beta^a = 0, \quad \alpha_a=a_a = const \Rightarrow$ \quad equations \eqref{44} have the form:
$$
n_{12}(a_1 + a_2) - n_{22}a_1-0, \quad n_{11}(a_1 + a_2) - n_{12}a_1-0.
$$
There are two independent solutions:

\quad

\noindent
${1}$ \quad $a_1 = 0,\quad n_{11} = n_{12} =0; $

\quad

\noindent
${2}$ \quad $a_2 = ca_1,\quad n_{12} = n_{11}(1+c), \quad n_{22} = n_{11}(1+c)^2. $

\quad

\noindent
\textbf{5.2}. \quad \textbf{GROUPS}\quad  $\textbf{G}_\textbf{3}\textbf{(V)} \quad\textbf{AND} \quad  \textbf{G}_\textbf{3}\textbf{(VI)}$

\quad

\noindent
Maxwell equations for the group $G_3(V)$ are obtained from Maxwell equations for the group $G_3(VI)$ with the zero value of the parameter $q$. Therefore, all solutions for the group $G_3(V)$ are special cases of solutions for the group $G_3(VI)$, and we will not consider the group $G_3(V)$ separately.
\quad

\noindent
Using structural equations and equations \eqref{6}, one can obtain the following  pair of sets of vectors \quad $\xi^\alpha_a,$ and $
\zeta^\alpha_a $:
\begin{equation}\label{54}
\zeta_a^\alpha =\delta_1^\alpha \delta_a^1 +\delta_2^\alpha \delta_a^2\exp(qu^3) +(\delta_3^\alpha - \delta_a^2 qu^1) \delta_a^3,
\quad \xi_{a}^{\alpha } =\delta _{1}^{\alpha } \delta _{a}^{1}\exp(-u^3) +\delta _{2}^{\alpha } \delta _{a}^{2}
 +(\delta_3^\alpha + \delta_a^2 qu^2) \delta_a^3.
\end{equation}

\noindent
The system of equations \eqref{23} has the form:
\begin{equation}\label{55}
qn_{11}\alpha_2 - n_{12}\alpha_1 =\dot{\beta}^2, \quad
\end{equation}
$$
q^2n_{12}\alpha_2 - q\alpha_1n_{22} =-\dot{\beta}^1.
$$
Let's consider the

\quad

\noindent
Option {\bf B} \quad $\beta^2 \ne 0, \quad \beta^3=0,\quad n_{12}=\frac{1}{\beta^2}(\dot{\alpha}_1 - n_{11}\beta^1), \quad n_{22}=\frac{1}{{\beta^2}^2}(\dot{\alpha}_2\beta^2 -\dot{\alpha}_1\beta^1 + n_{11}{\beta^1}^2). $

\quad

\noindent
The equations \eqref{55} will take the form:
\begin{equation}\label{56}
n_{11}(\beta^1\alpha_1 + q\beta^2 \alpha_2) =\alpha_1\dot{\alpha}_1 +\beta^2\dot{\beta}^2, \quad
\end{equation}
\begin{equation}\label{57}
q(q\dot{\alpha}_1\alpha_2 - \dot{\alpha}_2\alpha_1) + \dot{\beta}^1\beta^2 - q\dot{\beta}^2 \beta^1 =0.
\end{equation}
Let us find all solutions of these equations.

\quad

\noindent

Variant ${\bf B_1} $\quad Let $\quad\beta^1\alpha_1+q\beta^2 \alpha_2 \ne 0.$ Then:
\begin{equation}\label{58}
n_{11} =\frac{1}{\beta^1\alpha_1+q\beta^2 \alpha_2}(\alpha_1\dot{\alpha}_1+\beta^2\dot{\beta}^2), \quad
\end{equation}
There are the following independent solutions of the equation \eqref{57}.

\quad

\noindent
$1. \quad \alpha_1 =0, \quad \beta^1 = c{\beta^2}^q.$

\quad

\noindent
$2. \quad \alpha_2 =c_2{\alpha_1}^q, \quad \beta^1 = c_1{\beta^2}^q.$

\quad

\noindent
$3 \quad \alpha_1 = \beta^2(\frac{\dot{\gamma}}{\dot{\omega}})^{\frac{1}{q+1}},
\quad \alpha_2 = \omega (\beta^2)^q(\frac{\dot{\gamma}}{\dot{\omega}}),
\quad \beta_1 =q \gamma(\beta^2)^q. $

\quad

\noindent

Variant ${\bf B_2} $\quad Two more additional equations for the functions \quad $\alpha_p, \quad \beta^p $ appear:
$$
\quad \beta^1\alpha_1+q\beta^2 \alpha_2 = 0, \quad \beta^2\dot{\beta}^2 + \alpha_1\dot{\alpha}_1=0.
$$
Then the solutions of equation \eqref{57} have the following form:
$$
\beta^1 =-qb \sin\omega, \quad \alpha_2 = b \cos\omega, \quad \alpha_1 = a \cos\omega, \quad \beta^2 = a \sin\omega.
$$

\quad

\noindent
Option {\bf C}\quad Due to the existing symmetry of Maxwell equations with respect to the permutation \quad $1 \Leftrightarrow 2,$ \quad variant {\bf C} is a special case of the variant {\bf B}. That is why we will not consider it separately.

\quad

\noindent
Option {\bf D} \quad $\beta^a = 0, \quad \alpha_a = b_a = const$. Due to the symmetry mentioned above, there is a unique independent solution of the equations \eqref{32}, \eqref{74}:

\quad

\noindent
$$
a_2 = 0, \quad n_{12} = n_{22} = 0
$$.

\quad

\noindent
\textbf{5.3}. \quad \textbf{GROUP} $\textbf{G}_\textbf{3}\textbf{(VII)}$

\quad

\noindent
Let's choose vector fields of the canonical frame in the form:
\begin{equation} \label{59}
\zeta_{a}^{\alpha} =\delta _{a}^{1}(\delta_{1}^{\alpha } + \delta_{3}^{\alpha }(u^2 + 2u^3\cos a))     ) + \delta_{\alpha}^{2}\delta_{a}^{2} + \delta_{\alpha}^{3}\delta_{a}^{3}.
\end{equation}
Then from the equations \eqref{6} it follows:
\begin{equation}\label{60}
\xi_{a}^{\alpha } =\delta _{1}^{\alpha } \delta _{a}^{1} + \exp \gamma(\delta _{2}^{\alpha }(\delta_{a}^{2}\sin \sigma - \delta_{a}^{3}\cos \sigma ) + \delta _{3}^{\alpha }(\delta_{a}^{2}\sin(p+a) - \delta_{a}^{3}\cos(\sigma+a))).
\end{equation}
Here \quad $\sigma=u^0\sin a, \quad \gamma = u^0\cos a $.

\quad

\noindent
The system of equations \eqref{23} takes the form:
\begin{equation}\label{61}
n_{11}(q\alpha_2 - \alpha_1)-n_{12}\alpha_2 =\dot{\beta}^1 +q\dot{\beta}^2, \quad
\end{equation}
$$
n_{12}(q\alpha_2 - \alpha_1)-n_{22}\alpha_2 =\dot{\beta}^2.
$$
Consider

\quad

\noindent
Option {\bf B}

\quad

\noindent
Using equation \eqref{27}, one can represent equations \eqref{61} in the form:
\begin{equation}\label{62}
n_{11}(\beta^1\alpha_2 + \beta^2 (q \alpha_2-\alpha_1)) =\alpha_2\dot{\alpha}_1 +\beta^2(\dot{\beta}^1 +q\dot{\beta}^2), \quad
\end{equation}
\begin{equation}\label{63}
\dot{\alpha}_1 \alpha_1 +\dot{\alpha}_2\alpha_2+ \beta^1\dot{\beta}^1 + \beta^2\dot{\beta}^2 = q(\dot{\alpha}_1\alpha_2-\beta^1\dot{\beta}^2).
\end{equation}
Let's consider this system.

\quad

\noindent
${\bf B_1}\quad \beta^1\alpha_2 + \beta^2 (q \alpha_2-\alpha_1) \ne 0.$
From equation \eqref{62} it follows:
\begin{equation}\label{64}
n_{11} = \frac{1}{(\beta^1\alpha_2 + \beta^2 (q \alpha_2-\alpha_1))}(\alpha_2\dot{\alpha}_1 +\beta^2(\dot{\beta}^1 +q\dot{\beta}^2)).
\end{equation}
To solve the equation \eqref{63} let's introduce the function $\gamma$:
\begin{equation}\label{65}
2q\gamma = {\alpha_1}^2 + {\alpha_2}^2+{\beta_1}^2 + {\beta_2}^2 .
\end{equation}
Then equation \eqref{63} takes the form:
\begin{equation}\label{66}
\dot{\gamma} = (\dot{\alpha}_1\alpha_2-\dot{\beta}_2\beta_1).
\end{equation}
The derivatives of the functions  $\alpha_2, \beta^1$  are not presented in the equation \eqref{63}. By eliminating one of them  \quad( $\alpha_2$ or $\beta^1$)\quad from the system of equations \eqref{65},\eqref{65}, one obtains an equation for the remaining function. These functions will be expressed through the functions \quad $\alpha_1 ,\beta^2,  \gamma $ \quad and their first derivatives. Let's consider all possible options.

\quad

\noindent
1.\quad $\dot{\gamma} =0 \quad \Rightarrow \quad  \dot{\alpha}_1 = \dot{\beta}^2 =0 \quad \Rightarrow \quad \alpha_1 =a, \quad \beta^2 =b , \quad\alpha_2 =c \cos\omega, \quad \beta^1 = c \sin\omega.$

\quad

\noindent
2. \quad $\dot{\alpha}_1 \ne 0. $ \quad From equations \eqref{65}, \eqref{66} it follows:
\begin{equation}\label{67}
\alpha_2 = \frac{\varsigma}{\dot{\alpha_1}}\sqrt{\dot{\gamma} +\beta_1\dot{\beta^2}},
\quad \beta_1 = \frac{1}{(\dot{\alpha_1})^2 + (\dot{\beta^2})^2}(\dot{\gamma} \dot{\beta^2} +\Phi), \quad
\end{equation}
$$
\Phi^2 =((\dot{\gamma} \dot{\beta^2})^2+(\dot{\alpha_1}^2 + \dot{\beta^2}^2)({\dot{\alpha_1}}^2(2q\gamma -{\alpha_1}^2- {\beta^2}^2)-{\dot{\gamma}}^2).
$$

\quad

\noindent
Let's consider the

\quad

\noindent

Variant ${\bf B_2}$\quad $\beta^1\alpha_2 + \beta^2 (q \alpha_2-\alpha_1) = 0 \Rightarrow n_{11}$ \quad - arbitrary function of the variable $u^0$.

\quad

\noindent
The functions \quad $\alpha_a, \beta^a$ \quad obey the equations \eqref{63} and the system of equations:
\begin{equation}\label{68}
\alpha_2(\beta^1 + q \beta^2)=\alpha_1\beta^2, \quad
\beta^2(\dot{\beta}^1 +q\dot{\beta}^2) = \alpha_2\dot{\alpha}_1 .
\end{equation}
From equations \eqref{68} it follows:
\begin{equation}\label{69}
\alpha_2((\beta^1 + q \beta^2)(\dot{\beta^1} + q\dot{ \beta^2})+\alpha_1\dot{\alpha_1}) \Rightarrow (\beta^1 + q \beta^2)(\dot{\beta^1} + q\dot{ \beta^2})+\alpha_1\dot{\alpha_1}=0. \quad
\end{equation}
Therefore the solutions of the equations \eqref{63}, \eqref{68}, \eqref{69} have the form:

\quad

\noindent
$1 \quad \alpha_1 = a \sin\omega, \quad \alpha_2 = b\sin\omega, \quad \beta_1 = a(p-qb)\cos\omega, \quad \beta_2 = b \cos\omega, $

\quad

\noindent
$2 \quad \alpha_1=\alpha_2=0, \quad \beta^1 = const, \quad \beta^2 =const.$

\quad

\noindent
Let's consider the

\quad

\noindent
Option $ {\bf C} \quad \beta^2 = \beta_3 =0.$ \quad Equations \eqref{61} have the form:
\begin{equation}\label{70}
n_{11}(\alpha_1-q\alpha_2) + n_{12}a\alpha_2 = -\dot{\beta^1}, \quad n_{12}(\alpha_1-q\alpha_2) + n_{22}\alpha_2 = 0.
\end{equation}
Using the system of equations \eqref{28}:
$$
n_{1a} =\frac{\dot{\alpha_a}}{\beta^1},
$$
one can represents the system of equations \eqref{61} in the following form:
\begin{equation}\label{71}
n_{22}\beta^1\alpha_2 =\dot{\alpha}_2(q\alpha_2-\alpha_1), \quad
\end{equation}
\begin{equation}\label{72}
\dot{\alpha}_1 \alpha_1 +\dot{\alpha}_2\alpha_2+\beta^1\dot{\beta}^1 = q\dot{\alpha}_1\alpha_2.
\end{equation}
The equations \eqref{71}, \eqref{72} have next nonequivalent solutions:

\quad

\noindent
$1\quad \alpha_2\ne 0 \quad n_{22} =(q\alpha_2 -\alpha_1)\frac{\dot{\alpha_2}}{\beta_1\alpha_2}, $

\quad

\noindent
$ a) \quad \alpha_2 =\frac{\dot{\gamma}}{\dot{\alpha_1}}, \quad \beta_1 = \frac{\varsigma}{\dot{\alpha_1}}\sqrt{{\dot{\alpha_1}}^2(2q\gamma-{\alpha_1}^2) -{\dot{\gamma}}^2},$
%
%

\quad

\noindent
$b) \quad \alpha_1 = a,\quad \alpha_2 = p\cos\omega, \quad \beta^1=p \sin\omega.$

\quad

\noindent
$2 \quad \alpha_2 = 0 \Rightarrow n_{22} \quad $ is an arbitrary function, $\quad \alpha_1 = a\cos\omega, \quad \beta^1= a\sin\omega.$

\quad

\noindent
Let's consider the

\quad

\noindent
Option $ {\bf D}\quad \beta^a = 0, \quad \alpha_b = a_b = const$. \quad Equations \eqref{61} take the form:

\quad

\noindent
$$
n_{11}(qa_2 - a_1)- n_{12}a_2 = 0,\quad n_{12}(qa_2 - a_1)- n_{22}a_2 = 0.
$$
Since in the case of the group $G_3(VII)$, the indices 1, 2 enter into Maxwell equations and into the components of the metric tensor $g_{ij}$ non symmetrically, there are two nonequivalent solutions:

\quad

\noindent
$1 \quad $
$ a_1 = a_2q, \quad n_{12} = n_{22} = 0$,

\quad

\noindent
$2 \quad a_2 =0, \quad n_{12} = n_{11} = 0$.

\quad

\noindent
\section{SOLUTIONS OF MAXWELL EQUATIONS FOR UNSOLVABLE GROUPS}

\quad

\noindent
\textbf{6.1}. \quad \textbf{GROUP}  $\textbf{G}_{\textbf{3}}$\textbf{(VIII)}

\quad

\noindent
The set of generators of the group $G_3(VIII)$ for homogeneous Petrov spaces can be represented as:
\begin{equation}\label{73}
X_{1} =p_{3} \exp \left(-u^{2} \right), \quad X_{2} =p_{2}, \quad X_{3} =\left(p_{1} -u^{3} p_{2} +\frac{1}{2}\left({u^{3}}^{2} -\epsilon\right)p_3\right)\exp u^{2}\quad (\epsilon = 0, \pm 1).
\end{equation}
In the book \cite{10} the set of vectors \eqref{73} (more precisely: equivalent to the set \eqref{73} ) was used to classify homogeneous Petrov null spaces. At the same time, for solving the classification problem for homogeneous Petrov non-null spaces it have been used the set of vectors \eqref{74}:
\begin{equation}\label{74}{\rm X} _{1} =p_{3} \exp \left(-u^{2} \right), \quad{\rm X} _{2} =p_{2}, \quad {\rm X} _{3} =(p_{1} -u^{3} p_{2} +\frac{1}{2}{u^{3}}^{2}p_3) \exp u^{2},
\end{equation}
(see \cite{16}, p. 157). In the papers \cite{12}, \cite{77} the non-equivalence of the sets \eqref{73}, \eqref{74} has been proven.
Therefore, the classification of homogeneous non-null Petrov spaces with the group of motions $G_3(VIII)$ should be supplemented.
The easiest way to do this is to assume that the group operators \eqref{73} are defined by the vectors of the canonical frame  $\zeta^\alpha_a$. Then:
$$
g^{\alpha\beta} =\zeta^\alpha_a \zeta^\beta_b\eta^{ab},\quad Y_a=\zeta^\alpha_a p_a,
$$
where
\begin{equation}\label{75}
\zeta^\alpha_a =\delta_a^1 \delta^\alpha_3 \exp(-u^{2}) +\delta_a^2 \delta^\alpha_2  +\delta_a^3(\delta^\alpha_1 -u^{3} \delta^\alpha_2 +\frac{1}{2}({u^{3}}^{2}-\epsilon)\delta^\alpha_3) \exp u^{2}.
\end{equation}

\quad

\noindent
As already noted, the classification results using the canonical frame differ from the classification results using Petrov's method only in the choice of the holonomic coordinate system. Obviously, these results do not depend on the choice of values of the parameter $\epsilon$ (the choice of $\epsilon$ affects the form of the holonomic components of the metric tensor ($g_{\alpha\beta}$) and the holonomic components of the electromagnetic field vector $(A_\alpha $)).

The structure constants for \eqref{75} are of the form:
$$
C^1_{12} =C^{31}=1,\quad C^1_{23} =C^{13}=1,\quad C^2_{31} =C^{22}=-1\Rightarrow\quad f^1=\alpha_3, \quad f^2=-\alpha_2, \quad f^3=\alpha_1.
$$
Maxwell equations \eqref{22}-\eqref{23} take the form:
\begin{equation}\label{76}
\dot{\alpha_a}=\beta^b n_{ab}.
\end{equation}
\begin{equation}\label{77}
\alpha_1n_{33}-\alpha_2n_{23} +\alpha_3n_{13}=-\dot{\beta^1},
\end{equation}
$$
\alpha_1n_{23}-\alpha_2n_{22}+\alpha_3n_{12} =\dot{\beta^2},
$$
$$
\alpha_1n_{13} - \alpha_2n_{12} +\alpha_3n_{11}=-\dot{\beta^3}.
$$
These equations are symmetric with respect to the substitution \quad $\xi^\alpha_1 \Leftrightarrow\xi^\alpha_2$,
Therefore, given this symmetry, for a complete classification it is sufficient to consider options:

\quad

\noindent
$
{\bf B} \quad \beta^2 \ne 0, \quad {\bf C} \quad\beta^2 =0, \quad \beta^1 \ne 0.
$
\quad Option $\quad {\bf D}\quad(\beta^a =0$) \quad does not need to be considered, since for non-solvable groups \quad $G_3(VIII), G_3(IX)$ \quad it leads to a zero electromagnetic field.

\quad

\noindent
{\bf Option B}. $\quad \beta^2 \ne 0.\quad $
The equations \eqref{77} can be reduced to the form
\begin{equation}\label{78}
n_{12} = \frac{1}{\beta^2}(\dot{\alpha_1}-\beta^1 n_{11}-\beta^3 n_{13}),
\end{equation}

\begin{equation}\label{79}
n_{23} = \frac{1}{\beta^2}(\dot{\alpha_3}-\beta^1 n_{13}-\beta^3 n_{33}).
\end{equation}

\begin{equation}\label{80}
n_{22} = \frac{1}{{\beta^2}^2}(\dot{\alpha_2}\beta^2 - \dot{\alpha_3}\beta^3 -\dot{\alpha_1}\beta^1+{\beta^1}^2 n_{11}+2\beta^1{\beta^3}^2 n_{12}+ {\beta^3}^2 n_{33}).
\end{equation}
Using relations \eqref{78}-\eqref{80}, the remaining Maxwell equations can be represented as:

\begin{equation}\label{81}
n_{33}(\alpha_1\beta^2 + \alpha_2\beta^3 )+n_{13}(\alpha_3\beta^2 + \alpha_2\beta^1) =\alpha_2\dot{\alpha_3} -\beta^2\dot{\beta^1},
\end{equation}

\begin{equation}\label{82}
n_{13}(\alpha_1\beta^2 + \alpha_2\beta^3 )+n_{11}(\alpha_3\beta^2 + \alpha_2\beta^1) =\alpha_2\dot{\alpha_1}-\beta^2\dot{\beta^3},
\end{equation}

\begin{equation}\label{83}
{\beta^2}^2 +{\alpha_2}^2 -2 ({\alpha_1}{\alpha_3} + \beta^1 \beta^3) =c= const.
\end{equation}

\quad

Variant $ {\bf B_1.}$ \quad $(\alpha_1\beta^2 + \alpha_2\beta^3) \ne 0.$ \quad From equations \eqref{81}, \eqref{82} it follows:
\begin{equation}\label{84}
n_{13}=\frac{1}{\alpha_1\beta^2 + \alpha_2\beta^3 } (\beta^2\dot{\beta^3}+\alpha_2\dot{\alpha_1}- n_{11}(\alpha_3\beta^2 + \alpha_2\beta^1)),
\end{equation}
$$
n_{33}=\frac{1}{(\alpha_1\beta^2 + \alpha_2\beta^3 )^2}
 ((\beta^2\dot{\beta^1}+\alpha_2\dot{\alpha_3})(\alpha_1\beta^2 + \alpha_2\beta^3 )-(\beta^2\dot{\beta^3}+\alpha_2\dot{\alpha_1})(\alpha_3\beta^2 + \alpha_2\beta^1 )+ n_{11}(\alpha_3\beta^2 + \alpha_2\beta^1)^2).
$$
The solution to equation \eqref{83} can be written as
$$
\beta^2 =\varsigma\sqrt{ 2( {\alpha_1}{\alpha_3} + \beta^1 \beta^3)-{\alpha_2}^2 +c}.
$$

\quad

Variant  $ {\bf B_2.} \quad n_{33}$ is an arbitrary function \quad $\Rightarrow$
\begin{equation}\label{85}
\alpha_1\beta^2 + \alpha_2\beta^3 = 0.
\end{equation}
Let us designate:
$$
\alpha_1 = \gamma\alpha_2,\quad \beta^3=-\gamma\beta^2.
$$
Then solutions of equations \eqref{80} - \eqref{82}  have the form:
$$
n_{11}=\frac{\gamma}{\alpha_3\beta^2 + \alpha_2\beta^1}(\alpha_1\dot{\alpha_1} +\beta^3\dot{\beta^3}),\quad
n_{13}=\frac{\gamma}{\alpha_3\beta^2 + \alpha_2\beta^1}(\alpha_1\dot{\alpha_3}+\beta^3\dot{\beta^1}).
$$
The function $\gamma$ can been found from equations \eqref{82}, \eqref{85} and has the form:
$$
\gamma=\varsigma\sqrt{ \frac{c+2(\alpha_1\alpha_3 +\beta^1\beta^3)}{{\alpha_1}^2 +{\beta^3}^2} }
$$

\quad

Variant  $ {\bf B_3.} \quad n_{11}, n_{13}$ are arbitrary functions $\Rightarrow$ the functions $\alpha_a, \beta^a $ are determined from the system of equations:
\begin{equation}\label{86}
(\alpha_1\beta^2 + \alpha_2\beta^3) =\alpha_3\beta^2 + \alpha_2\beta^1 =0.
\end{equation}
\begin{equation}\label{87}
\beta^2\dot{\beta^1}+\alpha_2\dot{\alpha_3}=0, \quad
\beta^2\dot{\beta^3}+\alpha_2\dot{\alpha_1} =0,
\end{equation}
\begin{equation}\label{88}
\quad {\beta^2}^2 -{\alpha_2}^2 +2 {\alpha_1}{\alpha_3} - 2\beta^1 \beta^3 =c= const.
\end{equation}
Let us designate:
$$
\alpha_1=\gamma_3\alpha_2, \quad \alpha_3 = \gamma_1 \alpha_2, \quad \beta^1= -\gamma_1 \beta_2, \quad \beta^3= -\gamma_3 \beta_2.
$$
Then the equation \eqref{87} takes the form:
\begin{equation}\label{89}
\dot{\gamma_p}({\alpha_2}^2 +{\beta^2}^2 ) + \gamma_p(\alpha_2 \dot{\alpha_2} + \beta^2 \dot{\beta^2}) =0, \quad p=1, 3.
\end{equation}

\quad

\noindent
There are two nonequivalent solutions of the system of equations \eqref{89}:

\quad

\noindent
$1. \quad \gamma_q=0 \Rightarrow \alpha_q = \beta^q =0, \quad \alpha_2 = p\cos\omega, \quad \beta^2 = p\sin \omega $

\quad

\noindent
$2. \quad\gamma_p = c_p\gamma =\frac{c_p}{\sqrt{{\alpha_2}^2 +{\beta^2}^2 }}, \quad \alpha_1=c_3\gamma\alpha_2, \quad \alpha_3 = c_1 \gamma \alpha_2, \quad \beta^1= -c_1\gamma \beta_2, \quad \beta^3= -c_3\gamma \beta_2.$

\quad

\noindent
{\bf Option C}. \quad $\beta_2 = 0, \quad \beta_3 \ne 0. $

\quad

\noindent
The equations \eqref{77} can be reduced to the form:
\begin{equation}\label{90}
n_{13} = \frac{1}{\beta^3}(\dot{\alpha_1}-\beta^1 n_{11}),
\end{equation}

\begin{equation}\label{91}
n_{23} = \frac{1}{\beta^3}(\dot{\alpha_2}-\beta^1 n_{12}).
\end{equation}

\begin{equation}\label{92}
n_{33} = \frac{1}{{\beta^3}^2}(\dot{\alpha_3}\beta^3 - \dot{\alpha_1}\beta^1 + n_{11}{\beta^1}^2).
\end{equation}
Using relations \eqref{90}-\eqref{91}, the remaining equations from the Maxwell system of equations can be represented as:
\begin{equation}\label{93}
n_{11}(\alpha_1\beta^1 -\alpha_3\beta^3 ) + n_{12}\alpha_2\beta^3 =\beta^3\dot{\beta^3}+\alpha_1\dot{\alpha_1},
\end{equation}

\begin{equation}\label{94}
-n_{12}(\alpha_3\beta^3 - \alpha_1\beta^1 )+n_{22}\alpha_2\beta^3 =\alpha_1\dot{\alpha_2},
\end{equation}

\begin{equation}\label{95}
2({\beta^1}\beta^3 + {\alpha_1}{\alpha_3}) - {\alpha_2}^2 =c= const.
\end{equation}
Let's present all nonequivalent solutions.

\quad

Variant  $C_1 \quad \alpha_2 \ne 0. $ From equations \eqref{93}, \eqref{94} it follows:
\begin{equation}\label{96}
n_{22}=\frac{1}{({\alpha_2\beta^3})^2 }(\alpha_2\beta^3 (\alpha_1\dot{\alpha_2}-\beta^3\dot{\beta^2})+
(\alpha_3\beta^3 - \alpha_1\beta^1)(\alpha_1\dot{\alpha_1}+\beta^3\dot{\beta^3}) +n_{11}(\alpha_3\beta^3 - \alpha_1\beta^1)^2),
\end{equation}

\begin{equation}\label{97}
n_{12}=\frac{1}{\alpha_2\beta^3 }
(\alpha_1\dot{\alpha_1}+\beta^3\dot{\beta^3}+ n_{11}(\alpha_3\beta^3 + \alpha_1\beta^1)),
\end{equation}
$n_{11}$ is an arbitrary function. Solution of equation \eqref{95} can be present in the form:
$$
{\alpha_2}^2 =\varsigma\sqrt{2 ({\alpha_1}{\alpha_3}+{\beta^1}\beta^3) +c}.
$$

\quad

Variant $C_2$.$\quad \alpha_2 = 0,\quad \alpha_3\beta^3 - \alpha_1\beta^1\ne 0. $ Solution has the form:
$$
n_{11}=\frac{\beta^3\dot{\beta^3}+\alpha_1\dot{\alpha_1}}{\alpha_1\beta^1-\alpha_3\beta^3},
\quad n_{12}= 0, \quad \beta^1= \frac{c -\alpha_1\alpha_3}{\beta^3},
$$

\quad

Variant $C_3$.$\quad \alpha_1 = p\cos\omega, \quad \alpha_2 = 0,\quad \alpha_3 = c\cos\omega, \quad\beta^1 = c\sin \omega \quad \beta^2 =0, \quad \beta^3 = p\sin \omega $, \quad $n_{11}, n_{12}$ are arbitrary functions.

\quad

\noindent
\textbf{6.2}. \quad \textbf{GROUP} $\textbf{G}_{\textbf{3}}$\textbf{(IX)}

\quad

\noindent
The Killing vector fields $\xi^\alpha_a$ and the canonical frame vector fields $\zeta^\alpha_a$:
\begin{equation} \label{98}
\xi^\alpha_{a} = \delta^\alpha_1\delta^1_a + \delta^2_a(\delta^\alpha_{3} \cos u_{2} +\frac{\sin u^{2} }{\cos u^{3} }(\delta^\alpha_{1} +\delta^\alpha_{2} \sin u^{3})) + \delta^3_a\xi^\alpha_{2,2};
\end{equation}
\begin{equation} \label{99}
\zeta_{\left(a\right)}^{\alpha } = \delta^\alpha_1\delta^1_a +\delta^2_a((\delta^\alpha_1 \frac{\sin u^3}{\cos u^3}+ \delta^\alpha_2 \frac{1}{\cos u^3})\sin u^1
+\delta^\alpha_3\cos{u^1}) +\delta^3_a \zeta_{2,2}.\end{equation}
Hence: $$\quad f^a=\alpha_a, \quad C^{11}=C^{22}=C^{33}=1,$$
and Maxwell equations have the form:
\begin{equation}\label{100}
\alpha_1n_{a1}+\alpha_2n_{a2} +\alpha_3n_{a3}=-\dot{\beta^a},
\end{equation}
\begin{equation}\label{101}
\dot{\alpha_a}=\beta^b n_{ab}.
\end{equation}
Equations \eqref{100}-\eqref{101} are symmetric with respect to permutations of the vectors of the canonical frame.
Therefore, for a complete classification, it is sufficient to consider the variant $ \quad \beta^3 \ne 0. \quad $ In this case, the equations \eqref{22} can be reduced to the form:
\begin{equation}\label{102}
n_{13} = \frac{1}{\beta^3}(\dot{\alpha_1}-\beta^1 n_{11}-\beta^2 n_{12}),
\end{equation}

\begin{equation}\label{103}
n_{23} = \frac{1}{\beta^3}(\dot{\alpha_2}-\beta^1 n_{12}-\beta^2 n_{22}).
\end{equation}

\begin{equation}\label{104}
n_{33} = \frac{1}{{\beta^3}^2}(\dot{\alpha_3}\beta^3 - \dot{\alpha_1}\beta^1 -\dot{\alpha_2}\beta^2+{\beta^1}^2 n_{11}+2\beta^1{\beta^2} n_{12}+ {\beta^2}^2 n_{33}).
\end{equation}
Using relations \eqref{102}-\eqref{103}, the remaining Maxwell equations can be represented as:

\begin{equation}\label{105}
n_{11}(\alpha_3\beta^1-\alpha_1\beta^3 )+n_{12}(\alpha_3\beta^2-\alpha_2\beta^3) =\beta^3\dot{\beta^1}+\alpha_3\dot{\alpha_1},
\end{equation}

\begin{equation}\label{106}
n_{12}(\alpha_3\beta^1-\alpha_1\beta^3 )+ n_{22}(\alpha_3\beta^2-\alpha_2\beta^3) =\beta^3\dot{\beta^2}+\alpha_3\dot{\alpha_2},
\end{equation}

\begin{equation}\label{107}
{\alpha_1}^2 + {\alpha_2}^2+ {\alpha_3}^2 + {\beta_1}^2 + {\beta_2}^2 + {\beta_3}^2 =c^2= const.
\end{equation}

\quad

\noindent
In the case when $(\alpha_1\beta^3 - \alpha_3\beta^1) \ne 0,$ equations \eqref{104} - \eqref{105} allow one to represent the solution in the form:

\quad

Variant \textbf{1}.
\begin{equation}\label{108}
n_{12}=\frac{1}{\alpha_3\beta^1 -\alpha_1\beta^3 }(\beta^2\dot{\beta^3}+\alpha_3\dot{\alpha_2}+n_{22}( \alpha_2\beta^3 - \alpha_3\beta^2) ),
\end{equation}
\begin{equation}\label{109}
n_{11}=\frac{\beta^3\dot{\beta^1}+\alpha_3\dot{\alpha_1}}{\alpha_3\beta^1-\alpha_1\beta^3}
+\frac{(\alpha_2\beta^3-\alpha_3\beta^2 )}{(\alpha_1\beta^3 - \alpha_3\beta^1)^2}(\beta^2\dot{\beta^3}+\alpha_3\dot{\alpha_2}+n_{22}( \alpha_2\beta^3 -\alpha_3\beta^2)).
\end{equation}
Here  $n_{22}$ is an arbitrary function, $\alpha_a, \beta^1, \beta^3 $  obey the equation \eqref{106}, the solution of which can be represented as:
$$
\beta^2 = \varsigma\sqrt{{\alpha_1}^2 + {\alpha_2}^2 + {\alpha_3}^2 + {\beta^1}^2 + {\beta^3}^2 +c^2}.
$$

\quad

Variant \textbf{2}.\quad $(\alpha_2\beta^3 - \alpha_3\beta^2) \ne 0$,
\begin{equation}\label{110}
(\alpha_1\beta^3 - \alpha_3\beta^1) = 0.
\end{equation}
In this case, from equations {105}, {106} it follows that the functions  $n_{12}, n_{22}$ have the form:
$$
n_{12}=\frac{\beta^3\dot{\beta^1}+\alpha_2\dot{\alpha_1}}{\alpha_3\beta^2-\alpha_2\beta^3},
\quad n_{22}=\frac{\beta^3\dot{\beta^2}+\alpha_3\dot{\alpha_2}}{\alpha_3\beta^2-\alpha_2\beta^3},
$$
$n_{11}$ is an  arbitrary function. Let's denote \quad $\beta^1 = \gamma \beta^3, \quad \alpha_1 = \gamma\alpha_3$, \quad where $\gamma$ is an arbitrary function. \quad The solution of the equation \eqref{106}, \eqref{110} can be represented as:
$$
\beta^2= \varsigma\sqrt{{c^2-{\alpha_2}}^2 - ({\alpha_3}^2 +{\beta^3}^2)(1+{\gamma}^2)}
$$.

\quad

Variant \textbf{3}. \quad $n_{11}, n_{22}$ are arbitrary functions,
\begin{equation}\label{111}
\alpha_1\beta^3 - \alpha_3\beta^1) = (\alpha_2\beta^3 - \alpha_3\beta^2) = 0.
\end{equation}
Using the same notations as in the second variant, one obtains:
$$
\alpha_1 =\gamma\beta^1 \quad \alpha_2 =\gamma\beta^2, \quad \alpha_3 =\gamma\beta^3.
$$
The solution of the equation \eqref{106}, \eqref{111} can be represented as:
$$
\gamma= \varsigma\sqrt{\frac{c^2}{{\beta_1}^2 + {\beta_2}^2 + {\beta_3}^2}-1}
$$

\section{\bf LIST OF RESULTS}

This section presents all the solutions obtained. The results are presented in next manner. A separate subsection is devoted to each group. All non holonomic components of the metric tensor ($n_{ab}$) and electromagnetic potential ($\alpha_a$) are listed.The sets of Killing vectors $\xi_a^\alpha$ and canonical frame vectors $\zeta_a^\alpha$  are presented in the sections (4 - 6).

\quad

\noindent
\textbf{7.1}. \quad \textbf{SOLUTIONS FOR THE GROUP}  $\textbf{G}_{\textbf{3}}$\textbf{(I)}

\quad

\noindent
{\bf 1.}$ \quad n_{a3}= c\dot{\alpha_a}, \quad $ $\alpha_a, \quad n_{pq}$ \quad are arbitrary functions, \quad p, q = 1, 2.

\quad

\noindent
\textbf{7.2}. \quad \textbf{SOLUTIONS FOR THE GROUP}  $\textbf{G}_{\textbf{3}}$\textbf{(II)}

\quad

\noindent
{\bf 2.} \quad $n_{12}, n_{22}$ \quad are arbitrary functions,
$$ n_{11}=\frac{1}{\alpha_1}\dot{\beta}^1, \quad n_{13}=\frac{\dot{\alpha}_1\alpha_1+\beta^1\dot{\beta}_1}{\alpha_1}, \quad n_{23}=\dot{\alpha}_2-n_{12}\beta^1,\quad
n_{33}=\dot{\alpha}_3-\beta^1\frac{\dot{\alpha}_1\alpha_1+\beta^1\dot{\beta}_1}{\alpha_1}.
$$

\quad

\noindent
${\bf 3.} \quad n_{11}, n_{11}, n_{22}$\quad
are arbitrary functions,
$$\alpha^1=0,\quad \beta^1 =c, \quad n_{13}=-c n_{11}, \quad n_{23}=\dot{\alpha}_2 -c n_{12}, \quad n_{33}=\dot{\alpha}_3 +c^2
$$

\quad

\noindent
${\bf 4.}\quad n_{22}, n_{23}, n_{23} $ are arbitrary functions,
$$\alpha_1=a \sin\omega, \quad \beta^1= a\cos\omega \quad n_{11}=\dot{\omega}, \quad n_{12}=\frac{\dot{\alpha_2}}{a \cos\omega}\quad n_{13}=\frac{\dot{\alpha_3}}{a \cos\omega}.
$$

\quad

\noindent
\textbf{7.3}. \quad \textbf{SOLUTIONS FOR THE GROUP}  $\textbf{G}_{\textbf{3}}$\textbf{(III)}

\quad

\noindent
Option ${\bf B}$. \quad Functions $n_{2a}$ have the form:
$$
n_{12} = \frac{1}{\beta^2}(\dot{\alpha}_1-c n_{11}), \quad n_{22}= \frac{1}{{\beta^2}^2}(\dot{\alpha}_2\beta^2-\dot{\alpha}_1 c + c^2 n_{11}), \quad n_{23}=\frac{1}{\beta^2}(\dot{\alpha}_3 - n_{13}c).
$$
The rest functions are listed below ( $\dot{a}= \dot{a_a}=0$).

\quad

\noindent
${\bf5.}\quad n_{11}, n_{13}, n_{33}$ \quad are arbitrary functions $c=0,\quad \alpha_1=a \sin\omega, \quad \beta^2= a \cos\omega, \quad n_{12}=\dot{\omega}$.

\quad

\noindent
${\bf 6}.\quad  n_{13}, n_{33}$ \quad are arbitrary functions.
$$
 n_{11}=\frac{\dot{\alpha_1}\alpha_1+\beta^2\dot{\beta^2}}{c\alpha_1}, \quad n_{12}=-\frac{\dot{\beta^2}}{\alpha_1}.
$$

\quad

\noindent
${\bf 7} \quad n_{11}, n_{13},  n_{33}$. are arbitrary functions, \quad $\alpha_1 = 0, \quad \beta^2 =1.$
\quad

\quad

\noindent
Option ${\bf C}\quad n_{22}, n_{23}, n_{33}$ are arbitrary functions, $n_{1a}=\dot{\alpha_a}.$

\quad

\noindent
{$\bf 8.$}\quad $ \alpha_2 = const,$

\quad

\noindent
${\bf 9}.\quad \alpha_1 =0$.

\quad

\noindent
Option $\bf {D.}$\quad $\alpha_a=const$.

\quad

\noindent
$\bf {10.}$\quad $\alpha_1 =0.$

\quad

\noindent
$\bf {11.}$\quad $n_{12}=0.$

\quad

\noindent
\textbf{ 6.4} \quad \textbf{SOLUTIONS FOR THE GROUP} $\textbf{G}_\textbf{3}$\textbf{(IV)}

\quad

\noindent
Option ${\bf B}$ \quad The components $n_{2q}$ have the form:
$$
n_{12}=\frac{1}{\beta^2}(\dot{\alpha}_1 - n_{11}\beta^1), \quad n_{22}=\frac{1}{{\beta^2}^2}(\dot{\alpha}_2\beta^2 -\dot{\alpha}_1\beta^1 + n_{11}{\beta^1}^2).
\quad n_{23}=\frac{1}{\beta^2}(\dot{\alpha}_3 - n_{13}\beta^1),
$$
$n_{13}, n_{33}$ are arbitrary functions. The rest functions are listed below.

\quad

\noindent
${\bf 12.} \quad n_{11} = \frac{\dot{\beta}^2}{\alpha_2},\quad \beta_1 = \beta_2 (c -\ln \beta_2).$ \quad

\quad

\noindent
${\bf 13.} \quad n_{11}= \dot{\sigma} \quad \alpha_1 =\exp\sigma \cos\omega,\quad \beta^2 =\exp\sigma\sin\omega, \quad \alpha_2=\sigma \exp\sigma \cos\omega, \quad \beta^1 = -\sigma \exp\sigma\sin\omega.$

\quad

\noindent
$ {\bf 14}. \quad n_{11}= \frac{\dot{\sigma}}{\dot{\gamma_1}+\dot{\gamma_2}} \sqrt{(\dot{\sigma}+\dot{\gamma_1})(\dot{\gamma_2}-\dot{\sigma})},\quad\alpha_2 =\gamma_2 \alpha_1, \quad \alpha_1=\exp{\sigma} \sqrt{\frac{\dot{\sigma}+\dot{\gamma_1}}{\dot{\gamma_1}+\dot{\gamma_2}}},$
$$ \quad \beta^2 =\exp{\sigma} \sqrt{\frac{\dot{\gamma_2}-\dot{\sigma}}{\dot{\gamma_2}+\dot{\gamma_1}}}, \quad
\quad \beta^1 = \gamma_1 \beta^2. $$

\quad

\noindent
${\bf15.} \quad n_{11}$  is an arbitrary function,
$$
\alpha_1=(a_1 + a_2)\cos\omega, \beta^2=(a_1 + a_2)\sin\omega, \alpha_2=-a_2\cos\omega, \beta^1=-a_1\sin\omega.
$$

\quad

\noindent
Option {\bf C} \quad $ n_{1a}=\frac{\alpha_a}{\beta^1}. $

\quad

\noindent
${\bf16.} \quad n_{22}$ is an arbitrary function,\quad $n_{11} =0, \quad n_{12} =\dot{\omega},\quad \alpha_1 = 0,\quad \alpha_2=a\cos\omega, \quad \beta^1 = a \sin\omega.
$

\quad

\noindent
${\bf17.} \quad n_{11}$ is an arbitrary function,
$$
n_{22}=\frac{1}{\beta^1\alpha_1}(\dot{\alpha}_2\alpha_2+\dot{\beta}^1\beta^1 +\dot{ \alpha_2}\alpha_1).\quad \alpha_2 = \alpha_1(c+\ln{\alpha_1})
$$.

\quad

\noindent
Option {\bf D} \quad $\beta^a = 0, \quad \alpha_a=a_a = const.$

\quad

\noindent
${\bf18.} \quad  n_{11} = n_{12} =0,\quad a_1 = 0. $

\quad

\noindent
${\bf 19.} \quad n_{11}$ is an arbitrary function, \quad $n_{12} = n_{11}(1+c), \quad n_{22} = n_{11}(1+c)^2,\quad a_2 = ca_1. $

\quad

\noindent
\textbf{7.5} \quad \textbf{SOLUTIONS FOR THE GROUPS}  $\textbf{G}_{\textbf{3}}$ \textbf{(V)} \textbf{AND} $\textbf{G}_{\textbf{3}}$ \textbf{(VI)}

\quad

\noindent
Option ${\bf B}$. \quad Functions $n_{2a}$ have the form:
$$
n_{12} = \frac{1}{\beta^2}(\dot{\alpha}_1-\beta^1 n_{11}), \quad n_{22}= \frac{1}{{\beta^2}^2}(\dot{\alpha}_2\beta^2-\dot{\alpha}_1\beta^1  + {\beta^1}^2 n_{11}), \quad n_{23}=\frac{1}{\beta^2}(\dot{\alpha}_3 - n_{13}\beta^1 ).
$$
The rest functions are listed below.

\quad

\noindent

\quad

Variant ${\bf B_1}.\quad n_{11} =\frac{1}{\beta^1\alpha_1+q\beta^2 \alpha_2}(\alpha_1\dot{\alpha}_1+\beta^2\dot{\beta}^2)$

\quad

\noindent
${\bf 20.}$ \quad $\quad \alpha_1 =0, \quad \beta^1 = c({\beta^2})^q.$

\quad

\noindent
${\bf 21.}$ \quad$\quad \alpha_2 =c_2({\alpha_1})^q, \quad \beta^1 = c_1({\beta^2})^q.$

\quad

\noindent
${\bf 22.}$ \quad $\quad \alpha_1 = \beta^2(\frac{\dot{\gamma}}{\dot{\omega}})^{\frac{1}{q+1}},
\quad \alpha_2 = \omega (\beta^2)^q(\frac{\dot{\gamma}}{\dot{\omega}}),
\quad \beta_1 =q \gamma(\beta^2)^q. $

\quad

Variant ${\bf B_2}.\quad n_{11}$ is an arbitrary function.

\quad

\noindent
${\bf 23.}$ \quad $\beta^1 =-qa_1\sin\omega, \quad \alpha_2 = a_1\cos\omega, \quad \alpha_1 = a \cos\omega, \quad \beta^2 = a \sin\omega.$

\quad

\noindent
Option {\bf D}. \quad $\beta^a = 0, \quad \alpha_a=const$.

\quad

\noindent
${\bf 23.}$ \quad $ n_{12} = n_{22} = a_2 = 0.$

\quad

\noindent
\textbf{7.6} \quad \textbf{SOLUTIONS FOR THE GROUP}  $\textbf{G}_{\textbf{3}}\textbf{(VII)}$

\quad

\noindent
Option ${\bf B}$. \quad
Functions $n_{2a}$ have the form:
$$
n_{12} = \frac{1}{\beta^2}(\dot{\alpha}_1-\beta^1 n_{11}), \quad n_{22}= \frac{1}{{\beta^2}^2}(\dot{\alpha}_2\beta^2-\dot{\alpha}_1\beta^1  + {\beta^1}^2 n_{11}), \quad n_{23}=\frac{1}{\beta^2}(\dot{\alpha}_3 - n_{13}\beta^1 ).
$$
$n_{13}, n_{33}$  are arbitrary functions. The rest functions are listed below.

\quad

\noindent

Variant${\bf B_1}\quad n_{11} = \frac{1}{(\beta^1\alpha_2 + \beta^2 (q \alpha_2-\alpha_1))}(\alpha_2\dot{\alpha}_1 +\beta^2(\dot{\beta}^1 +q\dot{\beta}^2)).$

\quad

\noindent
${\bf 24} \quad \alpha_1 =a_1, \quad \beta^2 =a_2 , \quad\alpha_2 =a \cos\omega, \quad \beta^1 = a\sin\omega.$

\quad

\noindent
${\bf 25} \quad  \Phi^2 =((\dot{\gamma} \dot{\beta^2})^2+(\dot{\alpha_1}^2 + \dot{\beta^2}^2)({\dot{\alpha_1}}^2(2q\gamma -{\alpha_1}^2- {\beta^2}^2)-{\dot{\gamma}}^2),$
$$
\alpha_2 = \frac{\varsigma}{\dot{\alpha_1}}\sqrt{\dot{\gamma} +\beta_1\dot{\beta^2}},
\quad \beta_1 = \frac{1}{(\dot{\alpha_1})^2 + (\dot{\beta^2})^2}(\dot{\gamma} \dot{\beta^2} +\Phi).
$$

\quad

Variant ${\bf B_2}$ \quad $ n_{11}$ is an arbitrary function.

\quad

\noindent
${\bf 26} \quad \alpha_1 = a \sin\omega, \quad \alpha_2 = a_1\sin\omega, \quad \beta_1 = a \cos\omega(p-qb), \quad \beta_2 = a_1 \cos\omega.$

\quad

\noindent
${\bf 27}\quad \alpha_1=\alpha_2=0, \quad \beta^1 = const, \quad \beta^2 =const.$

\quad

\noindent
Option $ {\bf C}. \quad \beta^2=0, \quad n_{1a}= \frac{\dot{\alpha}}{\beta^1}.$

\quad

Variant $ {\bf C_1} \quad \alpha_2\ne 0, \quad n_{22} =(q\alpha_2 -\alpha_1)\frac{\dot{\alpha_2}}{\beta_1\alpha_2}, $

\quad

\noindent
${\bf 28} \quad \alpha_2 =\frac{\dot{\gamma}}{\dot{\alpha_1}}, \quad \beta_1 = \frac{\varsigma}{\dot{\alpha_1}}\sqrt{{\dot{\alpha_1}}^2(2q\gamma-{\alpha_1}^2) -{\dot{\gamma}}^2}.$
%
%

\quad

\noindent
${\bf 28}\quad \alpha_1 = a_1,\quad \alpha_2 = a\cos\omega, \quad \beta^1=a\sin\omega.$

\quad

Variant $ {\bf C_2} \quad \alpha_2=0, \quad n_{22} \quad $ is an arbitrary function,

\quad

\noindent
$ {\bf 29} \quad \alpha_1 = a\cos\omega, \quad \beta^1=a\sin\omega.$

\quad

\noindent

%
%
%
%

\quad

\noindent

Variant $ {\bf D}\quad \beta^a = 0, \quad \alpha_b = a_b = const$.

\quad

\noindent
$ {\bf 30}. \quad n_{12} = n_{22} = 0, \quad a_1 = a_2q, $.

\quad

\noindent
$ {\bf 31} \quad n_{12} = n_{11} = 0,\quad a_2 =0$.

\quad

\noindent
\textbf{7.7} \quad \textbf{SOLUTIONS FOR THE GROUP}  $\textbf{G}_{\textbf{3}}\textbf{(VIII)}$
%
%
%

\quad

\noindent
Option ${\bf B}$. \quad
Functions $n_{2a}$ have the form \eqref{34}:
$$
n_{12} = \frac{1}{\beta^2}(\dot{\alpha_1}-\beta^1 n_{11}-\beta^1 n_{13}),
\quad
n_{23} = \frac{1}{\beta^2}(\dot{\alpha_3}-\beta^1 n_{13}-\beta^3 n_{33}).
\quad
$$
$$
n_{22} = \frac{1}{{\beta^2}^2}(\dot{\alpha_2}\beta^2 - \dot{\alpha_3}\beta^3 -\dot{\alpha_1}\beta^1+{\beta^1}^2 n_{11}+2\beta^1{\beta^3}^2 n_{12}+ {\beta^3}^2 n_{33}).
$$

\quad

\noindent
$ {\bf 32.}$
$$
n_{13}=\frac{1}{\alpha_1\beta^2 + \alpha_2\beta^3 } (\beta^2\dot{\beta^3}+\alpha_2\dot{\alpha_1}- n_{11}(\alpha_3\beta^2 + \alpha_2\beta^1)), $$
$$
n_{33}=\frac{1}{(\alpha_1\beta^2 + \alpha_2\beta^3 )^2}
 ((\beta^2\dot{\beta^1}+\alpha_2\dot{\alpha_3})(\alpha_1\beta^2 + \alpha_2\beta^3 )-(\beta^2\dot{\beta^3}+\alpha_2\dot{\alpha_1})(\alpha_3\beta^2 + \alpha_2\beta^1 )+
 $$
 $$ n_{11}(\alpha_3\beta^2 + \alpha_2\beta^1)^2), \quad\beta^2 =\varsigma\sqrt{ 2( {\alpha_1}{\alpha_3} + \beta^1 \beta^3)-{\alpha_2}^2 +c}.
 $$

\quad

\noindent
$ {\bf 33.}$ \quad $n_{33}$ is an arbitrary function.
$$ \alpha_1 = \gamma\alpha_2,\quad \beta^3=-\gamma\beta^2. \quad
n_{11}=\frac{\gamma}{\alpha_3\beta^2 + \alpha_2\beta^1}(\alpha_1\dot{\alpha_1} +\beta^3\dot{\beta^3}),
$$
$$
n_{13}=\frac{\gamma}{\alpha_3\beta^2 + \alpha_2\beta^1}(\alpha_1\dot{\alpha_3}+\beta^3\dot{\beta^1}),
\quad \gamma=\varsigma\sqrt{ \frac{c+2(\alpha_1\alpha_3 +\beta^1\beta^3)}{{\alpha_1}^2 +{\beta^3}^2} }
$$

\quad

\noindent
$ {\bf 34.}\quad n_{11}, n_{13}$ are arbitrary functions, \quad $\alpha_1 = \alpha_3 =\beta^1 =\beta^3=0, \quad \alpha_2 = a\cos\omega, \quad \beta^2 = a\sin \omega.$

\quad

\noindent
$ {\bf 35.}\quad n_{11}, n_{13} $ are arbitrary functions,
$$ \quad\gamma_p = c_p\gamma =\frac{c_p}{\sqrt{{\alpha_2}^2 +{\beta^2}^2 }}, \quad \alpha_1=c_3\gamma\alpha_2, \quad \alpha_3 = c_1 \gamma \alpha_2, \quad \beta^1= -c_1\gamma \beta_2, \quad \beta^3= -c_3\gamma \beta_2.$$

\quad

\noindent
Option {\bf C}. \quad $\beta_2 = 0$,
$$
n_{13} = \frac{1}{\beta^3}(\dot{\alpha_1}-\beta^1 n_{11}),
\quad n_{23} = \frac{1}{\beta^3}(\dot{\alpha_2}-\beta^1 n_{12}).
\quad n_{33} = \frac{1}{{\beta^3}^2}(\dot{\alpha_3}\beta^3 - \dot{\alpha_1}\beta^1 + n_{11}{\beta^1}^2).
$$

\quad

\noindent
$ {\bf 36.}  \quad n_{11}$ is an arbitrary function,
$$
n_{22}=\frac{1}{({\alpha_2\beta^3})^2 }(\alpha_2\beta^3 (\alpha_1\dot{\alpha_2}-\beta^3\dot{\beta^2})+
(\alpha_3\beta^3 - \alpha_1\beta^1)(\alpha_1\dot{\alpha_1}+\beta^3\dot{\beta^3}) +n_{11}(\alpha_3\beta^3 - \alpha_1\beta^1)^2),
$$
$$
\quad n_{12}=\frac{1}{\alpha_2\beta^3 }
(\alpha_1\dot{\alpha_1}+\beta^3\dot{\beta^3}+ n_{11}(\alpha_3\beta^3 + \alpha_1\beta^1)),
\quad {\alpha_2}^2 =\varsigma\sqrt{2 ({\alpha_1}{\alpha_3}+{\beta^1}\beta^3) +c}.
$$

\quad

\noindent
$ {\bf 37.}\quad n_{22}$ is an arbitrary function, \quad
$ n_{11}=\frac{\beta^3\dot{\beta^3}+\alpha_1\dot{\alpha_1}}{\alpha_1\beta^1-\alpha_3\beta^3},
\quad n_{12}= 0, \quad \alpha_2 = 0,\quad \beta^1= \frac{c -\alpha_1\alpha_3}{\beta^3}.$

\quad

\noindent
$ {\bf 38.}$ \quad $n_{11}, n_{12}, n_{22}$ are arbitrary functions,
$$
\alpha_1 = a\cos\omega, \quad \alpha_2 = 0,\quad \alpha_3 = a_1\cos\omega, \quad\beta^1 = a_1\sin \omega \quad \beta^2 =0, \quad \beta^3 = a\sin \omega.
$$

\quad

\noindent
\textbf{7.8} \quad \textbf{SOLUTIONS FOR THE GROUP}  $\textbf{G}_{\textbf{3}}\textbf{(IX)}$
%
%
%

\quad

Functions $n_{3a}$ have the form:
$$
n_{13} = \frac{1}{\beta^3}(\dot{\alpha_1}-\beta^1 n_{11}-\beta^2 n_{12}),
\quad n_{23} = \frac{1}{\beta^3}(\dot{\alpha_2}-\beta^1 n_{12}-\beta^2 n_{22}).
$$
$$
n_{33} = \frac{1}{{\beta^3}^2}(\dot{\alpha_3}\beta^3 - \dot{\alpha_1}\beta^1 -\dot{\alpha_2}\beta^2+{\beta^1}^2 n_{11}+2\beta^1{\beta^2} n_{12}+ {\beta^2}^2 n_{33}).
$$
The rest functions are listed below.

\quad

\noindent
$\bf 39 \quad n_{22}$, is an arbitrary function.
$$
n_{12}=\frac{1}{\alpha_3\beta^1 -\alpha_1\beta^3 }(\beta^2\dot{\beta^3}+\alpha_3\dot{\alpha_2}+n_{22}( \alpha_2\beta^3 - \alpha_3\beta^2) ),
$$
$$
n_{11}=\frac{\beta^3\dot{\beta^1}+\alpha_3\dot{\alpha_1}}{\alpha_3\beta^1-\alpha_1\beta^3}
+\frac{(\alpha_2\beta^3-\alpha_3\beta^2 )}{(\alpha_1\beta^3 - \alpha_3\beta^1)^2}(\beta^2\dot{\beta^3}+\alpha_3\dot{\alpha_2}+n_{22}( \alpha_2\beta^3 -\alpha_3\beta^2)),
$$
$$
\beta^2 = \varsigma\sqrt{c^2-{\alpha_1}^2 - {\alpha_2}^2 - {\alpha_3}^2 - {\beta^1}^2 - {\beta^3}^2}.
$$

\quad

\noindent
${\bf 39} \quad n_{11}$ is an arbitrary function,
$$
n_{12}=\frac{\beta^3\dot{\beta^1}+\alpha_2\dot{\alpha_1}}{\alpha_3\beta^2-\alpha_2\beta^3},
\quad n_{22}=\frac{\beta^3\dot{\beta^2}+\alpha_3\dot{\alpha_2}}{\alpha_3\beta^2-\alpha_2\beta^3},
$$
$$\beta^1 = \gamma \beta^3, \quad \alpha_1 = \gamma\alpha_3,\quad
\beta^2= \varsigma\sqrt{{c^2-{\alpha_2}}^2 - ({\alpha_3}^2 +{\beta^3}^2)(1+{\gamma}^2)}.
$$

\quad

\noindent
${\bf 40} \quad n_{11}, \quad n_{22}$ \quad are arbitrary functions,
$$
\alpha_1 =\gamma\beta^1 \quad \alpha_2 =\gamma\beta^2, \quad \alpha_3 =\gamma\beta^3. \quad \gamma= \varsigma\sqrt{\frac{c^2}{{\beta_1}^2 + {\beta_2}^2 + {\beta_3}^2}-1}.
$$

\quad

\noindent

\section{COLCLUTION}

Results obtaining in this paper can be used to solve the problem of classifying exact solutions of the vacuum Einstein-Maxwell equations for homogeneous Petrov non-null spaces. For Stackel spaces this classification problem has been solved to a large extent (including in our papers, see, for example, \cite{19} and the literature cited there). However, for Petrov spaces this problem is at an early stage of solution. A complete classification has been carried out only for spaces with an additive group of motions $ G_3(I)$ (see \cite{40}-\cite{41}).The classification of admissible electromagnetic fields and exact solutions of Maxwell vacuum equations and the presentation of the results in a form convenient for use are the initial stage of the classification of electrovacuum spaces associated with homogeneous Petrov spaces.

\quad

%
%
%

\quad



\quad

\



\begin{thebibliography}{99}

\bibitem{1}
Stackel. P. Uber die intagration der Hamiltonschen differentialechung mittels separation der variablen.  {\em Math. Ann.}
{\bf 1897}. {\em 49}, (145-147 pp.);

\bibitem{2}
 Eisenhart L.P. Separable systems of stackel. {\em Ann.Math.} {\bf 1934}, {\em 35}, (284-305 pp).;

\bibitem{3}
Levi-Civita T. Sulla Integraziome Della Equazione Di Hamilton-Jacobi Per Separazione Di Variabili. {\em Math.Ann.} {\bf 1904} {\em 59}, (383-397 pp.);

\bibitem{4}
Jarov-Jrovoy M.S. Integration of Hamilton-Jacobi equation by complete separation of variables method. {\em J.Appl.Math.Mech.}. {\bf 1963}, {\em27}, {\em No 6},  (173-219 pp).org/10.1016/0021-8928(63)90122-9;

\bibitem{5}
Carter B. New family of Einstein spaces. {\em Phys.Lett.}
{\bf 1968}, {\em A.25}, {\em No 9} (399-400pp.), doi.org/10.1016/0375-9601(68)90240-5;

\bibitem{6}
 Shapovalov V.N., Symmetry and separation of variables in the Hamilton-Jacobi equation. {\em Sov. Phys.J.}.  {\bf 1978}, {\em 21}, (1124-1132pp.). doi: 10.1007/BF00894560;

\bibitem{7}
 Shapovalov V.N., Stackel`s spaces. {\em Sib. Math. J.} {\bf 1979}, {\em 20}, (1117-1130 pp.), doi: org/10.1007/BF00971844;

\bibitem{8}
Shapovalov V.N., Symmetry of motion equations of free particle  in riemannian space. {\em Sov. Phys.J.} {\bf 1975}, {\em 18}, (1650-1654pp.), doi.org/10.1007/BF00892779;

\bibitem{9}
Miller W.  Symmetry And Separation Of Variables. {\em Cambridge University Press:Cambridge}. {\bf1984}, (318 p.p.);

\bibitem{10}
 Petrov, A.Z. \emph{Einstein Spaces}; {Pergamon press, Original owner: University of California}
 1969; 494p;  {Russian original published by Nauka, Moscow, 1961.}

\bibitem{11}	
Petrov, A.Z. \emph{New Methods in General Relativity}; {Nauka: Moscow}, {Russia,}  {1966}; 496p. (In Russian)

\bibitem{12}

Obukhov, V.V.  Classification of Petrov homogeneous spaces. {\em Symmetry.}
{\bf 2024}, 16, 1385 (15p.). doi.org/10.3390/sym16101385.

\bibitem{13}
Shapovalov V.N., Eckle G.G., Separation of Variables in the Dirac Equation. {\em Sov. Phys. J.} {\bf 1973}, 16, 6 (818-823pp.), doi.org/10.1007/BF00895697;

\bibitem{14}
Shapovalov V.N., Symmetry and separation of variables in a linear second-order differential equation. I, II. {\em Sov. Phys.J.}, {\bf 1978} {21}, (645-650, 693-695 pp.) doi.org/10.1007/BF00890983;

\bibitem{15}
Bagrov, A. G. Meshkov, V. N. Shapovalov, A. V. Shapovalov. Separation of variables in the Klein-Gordon equations I. {\em Sov. Phys.J} {\bf1973}, {\em16}, (1533-1538 pp.). doi.org/ 10.1007/BF00889957;

\bibitem{16}
Bagrov, V.G., Meshkov, A.G., Shapovalov, V.N. et al. Separation of variables in the Klein-Gordon equations. II. \emph{Soviet Physics Journal} \textbf{1973}, 16, 1659–1665. https://doi.org/10.1007/BF00893656

\bibitem{17}
 Bagrov V. G., Meshkov A. G., Shapovalov V. N., Shapovalov A. V. Separation of variables in the Klein-Gordon equations $III$. {\em Sov. Phys. J.} {\bf 1974}, {\em17}, (812-815 pp.) doi:org/10.1007/BF00890216;


\bibitem{18}
Obukhov V V and Osetrin K E and Shapovalov, A V.
Comments on the article M.O. Katanaev, Complete separation of variables in the geodesic Hamilton-Jacobi equation in four dimensions. {\em Physica Scripta}, {\bf2023}, 98, 104001. doi:10.1088/1402-4896/ad4182.
\bibitem{19}
 Obukhov V.V. Separation of variables in Hamilton-Jacobi and Klein-Gordon-Fock equations for a charged test particle in the Stackel spaces of type (1.1). {\em Int. J. Geom. Meth. Mod. Phys}. {\bf 2021}, {\bf18}, {03}, (2150036). doi:10.1142/S0219887821500365

 \bibitem{20}
Shapovalov A.V., Shirokov I.V. Noncommutative integration method for linear partial differential equations. functional algebras and dimensional reduction. {\em Theoret. And Math. Phys.} {\bf 1996}, {106:1}, 1-10. doi.org/10.4213/tmf1093.

\bibitem{21}
Breev A.I., Shapovalov A.V. Non-commutative integration of the Dirac equation in homogeneous spaces. {\em Symmetry},{\bf 2020}, {\em 12}, 1867. doi.org/10.3390/sym12111867;

\bibitem{22}
Magazev A.A. , Integrating Klein-Gordon-Fock equations in an extremal electromagnetic
field on Lie groups. {\em Theor.and Math.Phys.} {\bf 2012} 173:3, 1654-1667, doi: 10.1007/s11232-
012-0139-x, arxiv.org/abs/1406.5698;
\bibitem{23}
Magazev A. A. , Shirokov I. V. , Yurevich Yu. A.  Integrable magnetic geodesic flows on Lie groups, {\bf 2008}, TMF, 156:2, 189-206; {\em Theoret. and Math. Phys}. {\bf 2008} 156:2, 1127-1141. doi.org/10.4213/tmf624;

\bibitem{24}
Magazev A.A., Constructing a Complete Integral of the Hamilton-Jacobi Equation on
Pseudo-Riemannian Spaces with Simply Transitive Groups of Motions. {\em Math. Phys. Anal
Geom.} 24, 11, {\bf 2021}. doi.org/10.1007/s11040-021-09385-


\bibitem{25}
Obukhov V. V. Hamilton-Jacobi and Klein-Gordon-Fock equations for a charged test
particle in space-time with simply transitive four-parameter groups of motions. {\em J. Math. Phys.}, {\bf 2023} 64, 093507; doi: 10.1063/5.0158054;



\bibitem{26}
Obukhov V.V. Algebra of the symmetry operators of the Klein-Gordon-Fock equation for the case when groups of motions $G_3$ act transitively on null subsurfaces of spacetime.
{\em Symmetry}. {\bf 2022}, 14, (346). doi.org/10.3390/sym14020346;

\bibitem{27}
Schwarzschild, K. Uber das Gravitationsfeld eines Masenpunktes nach der
Einsteinschen Theorie.{\em Sitz. Preuss. Akad. Wiss.}, {\bf 1916}, Seite 189-196;

\bibitem{28}
Schwarzschild, K.  Uber das Gravitationsfeldeiner Kugel aus inkompressibler Flussigkeit nach der Einsteinschen
Theorie. {\em Sitz. Preuss. Akad. Wiss.}, {\bf 1916}, 424-434;

\bibitem{29}
Kerr, R.P.. Gravitational field of a spinning mass as an example of algebraically
special metrics.  Phys. Rev. Lett. {\bf 1963}, 11, 237, doi.org/10.1103/PhysRevLett.11.237;

\bibitem{30}
Newman,E. Tamburino L. and Unti T.,Empty space generalization of the Schwarzschild metric,J. Math. Phys. {\bf1963}, 915,doi:10.1063/1.1704018

\bibitem{31}
Reissner, H. Uber die Eigengravitation des elektrischen Feldes nach der Einsteinschen Theorie., Annalen der Physik,
{\bf 1916}, 355, 9, (106-120), doi:org/10.1002/andp.19163550905

\bibitem{32}
Friedmann, A. Uber die Moglichkeit einer Welt mit konstanter negativer Krummung des Raumes. Z. Physik, {\bf 1924}, 21, 326-332 . doi.org/10.1007/BF01328280

\bibitem{33}
Taub, A. H. Empty Space-Times Admitting a Three Parameter Group of Motions. Annals of Mathematics. {\bf
    1951}, 53, no. 3: 472-490. doi.org/10.2307/1969567.

\bibitem{34}
Stephani H., Kramer D., Mac Callum  M., Hoenselaers C., and Herlt E. Exact Solutions of Einstein's Field Equations. Second Edition.{\em Cambridge University Press. Cambridge.} {\bf 2003}, (732 pp.) ISBN 0521461367.doi: https://doi.org/10.1017/CBO9780511535185)

\bibitem{35}
Chong Z.W., Gibbons G,W. and Pope C.N. Separability and Killing tensors in Kerr-Taub-Nut-De Sitter metrics in higher dimensions. {\em Phys. Lett. B.} {\bf 2005}, {\em 609}, (124-132pp.), doi: 10.1016/j.physletb.2004.07.066;

\bibitem{36}
Vasudevan M. , Stevens K.A. And Page D.N. Separability Of The Hamilton-Jacobi And Klein-Gordon Equations In Kerr-De Sitter Metrics, {\em Class. And Quant. Grav.} {\bf 2005}, {22}, (339-352pp.), doi: 10.1088/0264-9381/22/2/007;

\bibitem{37}
 Frolov. P. , Kyoto U., Krtous P., Kubiznak D. Separation of variables in maxwell equations in Plebanski-Demianski spacetime {\em Phys.Rev.} {\bf 2018}, {\em D 97}, {\em No.10}, 101701 (6 pp.), doi:10.1103/PhysRevD.97.101701;

\bibitem{38}
Akbar M.M., MacCallum M.A.H.  Static axisymmetric Einstein equations in vacuum: Symmetry, new solutions, and Ricci solitons.  {\bf 2015},  {\em  Physical Review D}, 92, 6, id.063017, 10.1103/PhysRevD.92.063017;

\bibitem{39}
Kumaran Y., Ovgun A. Deflection angle and shadow of the
reissner-nordstrom black hole with higher-order magnetic correction in
einstein-nonlinear-maxwell fields. {\em Symmetry}, {\bf 2022}, 14, 2054. doi.org/10.3390/sym14102054;

\bibitem{391}
Brevik I.,Obukhov V.V. , Timoshkin A.V.  Cosmological Models Coupled with Dark Matter in a Dissipative Universe. {\em Astrophys.Space Sci}. {\bf 2015}, 359, 1, 11, e-Print: 1507.05216 [gr-qc]
doi: 10.1007/s10509-015-2451-z;

\bibitem{40}
Obukhov V.V. Classification of Einstein spaces with Stackel metric of type (3.0), {\em International Journal of Geometric Methods in Modern Physics}. {\bf 2025},  2550177 (14 pages).  doi: 10.1142/S0219887825501774;

\bibitem{41}
Obukhov V.V. Classification of the non-null electrovacuum solution of Einstein–Maxwell equations with three-parameter abelian group of motions,  {\em Ann. of Phys.}, {\bf 2024}, 470, 169816. doi.org/10.1016/j.aop.2024.169816;


\bibitem{401}
Valenzuela M. Hamiltonian formalism of Bianchi models with cosmological constant. \emph{Revista de Investigacion de Fisica,}  \textbf{2025}, 28, 1. doi: 10.15381/rif.v28i1.27302.

\bibitem{402}
Saez J. A. et al. Spatially-homogeneous cosmologies  \emph{Class. Quantum Grav.} \textbf{2024}, 41, 205013. doi: 10.1088/1361-6382/ad7664

\bibitem{42}
Francisco Astorga, J Felix Salazar and Thomas Zannias.
On the integrability of the geodesic flow on a Friedmann-Robertson-Walker spacetime,  {\em Physica Scripta}, {\bf 2022},. {\bf 93}, {\em 8}, 93 085205,.
doi. 10.1088/1402-4896/aacd44;

\bibitem{43}
Dappiaggi Claudio, Benito A. Juarez-Aubry, Alessio Marta Ground. State for the Klein-Gordon field in anti-de Sitter spacetime with dynamical Wentzell boundary conditions. {\em Phys. Rev. D}. {\bf2022}, 105017, 105. doi.org/10.1103/PhysRevD.105.105017;

\bibitem{44}
Nojiri S., Odintsov S.D., and Oikonomou V.K. Modified gravity theories on a nutshell: Inflation, bounce and late-time evolution.  {\em Phys. Rept.}, {\bf2017}, (692pp.), doi:10.1016/j.physrep.2017.06.001;

\bibitem{45}
Bamba  K., Capozziello S., Nojiri  S. and Odintsov S.D.
Dark energy cosmology: the equivalent description via different theoretical models and cosmography tests. {\em Astrophys. Space Sci.},{\bf 2012}, {\em342}, (155pp.), doi: 10.1007/s10509-012-1181-8;

\bibitem{46}
Capozziello S., De Laurentis M., Odintsov D. Hamiltonian dynamics and Noether symmetries in extended gravity cosmology.  {\em Eur.Phys.J.} {\bf 2012}, {\em C72}, 2068 (22 pp.), doi: 10.1140/epjc/s10052-012-2068-0;

\bibitem{47}
Kibaroglu S. Odintsov S.D. and Paul T.
Cosmology of unimodular Born Infeld-fR gravity, {\em Phys. Dark Univ.} {\bf 2024}, 44, 101445. doi:10.1016/j.dark.2024.101445
[arXiv:2402.08951 [gr-qc]];

\bibitem{48}
Kibaroglu S. Cosmological application of the Maxwell gravity. {\em Int. J. Mod. Phys. D.} {\bf 2023}, 32, 10, 2350069. doi:10.1142/S0218271823500694
[arXiv:2208.06641 [hep-th]];

\bibitem{50}
Odintsov S.D. Modified gravity: the game of theories?. {\em Russian Physics Journal}, {\bf 2025}, 68(1), 108-112. doi: 10.1007/s11182-025-03408-0;

\bibitem{51}
Kibaroglu S.~Cebecioglu O.~ and Saban A.
Gauging the Maxwell Extended and Algebras, {\em Symmetry}, {\bf 2023}, 15, 2, 464. doi:10.3390/sym15020464.


\bibitem{53}
Brevik I, V.V. Obukhov, A.V. Timoshkin. Quasi-Rip and Pseudo-Rip Universes Induced by the Fluid Inhomogeneous Equation of State.  {\em Astrophys.Space Sci}. {\bf 2022}, 344275-279. doi: 10.1007/s10509-012-1328-7

\bibitem{54}
Oktay Cebecioglu, Ahmet Saban, Salih Kibaroglu, Maxwell extension of f(R) gravity
{\em Eur. Phys. J. C}, {\bf 2023}, 83:95. doi.org/10.1140/epjc/s10052-023-11185-8

\bibitem{55}
Paliathanasis A. Kantowski-Sachs cosmology in scalar-torsion theory, {\em Eur. Phys. J. C,} {\bf 2023}, 83 no.3, 213.
doi:10.1140/epjc/s10052-023-11342-z
[arXiv:2302.09608 [gr-qc]].

\bibitem{56}
Medine Ildesa, Metin Arikb. Analytic solutions of scalar field cosmology, mathematicalstructures for early inflation and late time accelerated expansion. {\em Eur. Phys. J. C}. {\bf2023}, 83:167, doi.org/10.1140/epjc/s10052-023-11273-9

\bibitem{57}
Osetrin K., Osetrin E. Shapovalov Wave-Like Spacetimes. {\em Symmetry}, {\bf 2020}, 12, 1372. doi.org/10.3390/sym12081372

\bibitem{58}
Osetrin K.E., Epp V.Y., Filippov A.E.
Exact Model of Gravitational Waves and Pure Radiation.
{\em Symmetry}, {\bf 2024}, 16, 11, 1456. doi.org/10.3390/sym16111456

\bibitem{59}
Osetrin, K.E.;  Epp, V.Y.;  Chervon, S.V.
Propagation of light and retarded time of radiation in a strong gravitational wave,
{\em Annals of Physics}, {\bf 2024}, 462, 169619. doi.org/10.1016/j.aop.2024.169619

\bibitem{60}
Osetrin, K.E., Epp, V.Y., Guselnikova, U.A.
Gravitational wave and pure radiation in the Bianchi IV universe.
{\em Russ Phys J}, {\bf 2024}, 68, 526–531. doi.org/10.1007/s11182-025-03460-w

\bibitem{61}
Osetrin K. Osetrin E. Osetrina E.
Deviation of Geodesics, Particle Trajectories and the Propagation of Radiation in Gravitational Waves in Shapovalov Type $III$ Wave Spacetimes.
{\em Symmetry}, {\bf 2023}, 15, 7, 1455. doi.org/10.3390/sym15071455


\bibitem{62}
Camci U. Noether Symmetry Analysis of the Klein–Gordon and Wave Equations in Bianchi $I$ Spacetime. {\em Symmetry}, {\bf 2024}, 16, 115. doi.org/10.3390/sym16010115

\bibitem{63}
Magazev A.~A.~ and Boldyreva M.~N.
Schrodinger Equations in Electromagnetic Fields: Symmetries and Noncommutative Integration,{\em Symmetry.} {\bf 2021}, \textbf{13}, no.8, 1527. doi:10.3390/sym13081527

\bibitem{64}
Breev A.~I.~, Shirokov I.~V.~ and Magazev A.~A.
Vacuum polarization of a scalar field on lie groups and homogeneous spaces,
{\em Theor. Math. Phys.} {\bf 2023}, \textbf{167}, 468-483. doi:10.1007/s11232-011-0035-9

\bibitem{65}
Komrakov B. B.,  {\em J. Math.} {\bf 2001} 8, 33-165.

\bibitem{66}
Calvaruso G., and Fino A.  Four-dimensional pseudo-Riemannian homogeneous Ricci solitons. International. {\em Journal of Geometric Methods in Modern Physics}, {\bf 2015}, 12(05), 1550056. doi.org/10.48550/arXiv.1111.6384;

\bibitem{67}
Calvaruso G., and Amirhesam Zaeim. A complete classification of Ricci and Yamabe solitons of non-reductive homogeneous 4-spaces. {\em Journal of Geometry and Physics}, {\bf 2015}, 80, 15-25. doi.org/ 10.1016/j.geomphys.2014.02.007;

\bibitem{68}
Ugur Camci. Noether Symmetry Analysis of the Klein-Gordon and Wave Equations in Bianchi I Spacetime. {\em Symmetry}, {\bf 2024}, 16(1), 115; doi.org/10.3390/sym16010115;

\bibitem{69}
Ghezelbash A. M. Bianchi $IX$ geometry and the Einstein-Maxwell theory {\em Class. Quantum Grav.} 39 {\bf 2022} 075012 (36pp) doi.org/10.1088/1361-6382/ac504e;

\bibitem{70}
Osetrin E.  Osetrin K.  Filippov A.
 Plane Gravitational Waves in Spatially-Homogeneous Models of type-(3.1) Stackel Spaces.
 {\em Russian Physics Journal} {\bf 2019}, {\em 62}, ~292-301.
doi.org/10.1007/s11182-019-01711-1;

\bibitem{71}
Osetrin E.K., Osetrin K.E. Filippov A.E. Spatially Homogeneous Conformally Stackel Spaces of Type (3.1). {\bf 2020}, {\em  Russ Phys J.} 63, 403–409. doi.org/10.1007/s11182-020-02050-2;

\bibitem{72}
Osetrin E.~Osetrin K.~, and Filippov A.~ Spatially Homogeneous Models Stackel Spaces of Type (2.1), {\em Russian Physics Journal}, {\bf 2020}, 419, 63. doi.org/10.1007/s11182-020-02051-1;

\bibitem{721}
Camci U. Noether Symmetry Analysis of the Klein–Gordon and Wave Equations in Bianchi I Spacetime. \emph{Symmetry,} \textbf{2024}, 16, 115. https://doi.org/10.3390/sym16010115;

\bibitem{722} Magazev A.A., Shirokov I.V. The Structure of Differential Invariants for a Free Symmetry Group Action. \emph{Russ Math}. \textbf{2023}, 67, 26–33.doi.org/10.3103/S1066369X23060051;

\bibitem{723} Shair-a-Yazdan, Muhammad Jawed Iqbal, and Mark Hickman. Anisotropic models in magnetized Bianchi Type-III spacetimes via noether symmetries.  \emph{Modern Physics Letters A.}  \textbf{2025},  (40.07n08): 2550003. doi: 10.1142/S0217732325500038;

\bibitem{73}
Obukhov V.V. Maxwell Equations in Homogeneous Spaces for Admissible Electromagnetic Fields,
{\em Universe}  {\bf 2022} no.4, 245 doi:10.3390/universe8040245;

\bibitem{74}
Obukhov V.V.  Maxwell Equations in Homogeneous Spaces with Solvable Groups of Motions. {\em Symmetry}. {\em 14}, 2595, (2022).doi.org/10.3390/sym14122595;

\bibitem{75}
 V. V. Obukhov. Exact Solutions of Maxwell Equations in Homogeneous Spaces with the Group of Motions $G_3(IX)$. Axioms.{\bf 2023}.{\em 12}, {\em 135}, doi.org/10.3390/axioms12020135;

\bibitem{76}
Obukhov V.V., Exact Solutions of Maxwell Equations in
Homogeneous Spaces with the
Group of Motions $G_3(VIII)$. {\em Symmetry.} {\em 15}, 648, {\bf2023}. doi.org/10.3390/sym15030648;

\bibitem{77}
Obukhov V.V. and Kartashov D.V. {\em Non-null homogeneous Petrov type $VIII$ space-time by Bianchi classification.}
{\em Russ. Phys. J}. {\bf 2024}, 67, 11, 1913-1917. doi:10.1007/s11182-024-03327-6;

\bibitem{78}
Landau, L.D.; Lifshits, E.M. Theoretical Physics, Field Theory, 7th ed.; {\em Science, C.,
Ed.; Nauka: Moskow, Russia,} 1988; Volume II, 512p. ISBN 5-02-014420-7.

\end{thebibliography}
\end{document}